# Relativistic Effects in Atom and Neutron Interferometry
# And the Differences Between Them

by

Daniel M. Greenberger[1]
*City College of New York, New York, NY 10031 USA*

Wolfgang P. Schleich[2]
*Institut für Quantenphysik, Center for Integrated Quantum Science and Technology (IQST), Universität Ulm, Albert Einstein Allee 11, D-89069 Ulm, Germany*

and

Ernst M. Rasel[3]
*Institut für Quantenoptik, Leibniz Universität Hannover, Welfengarten 1, D-30167 Hannover, Germany*

## Abstract

In recent years there has been an enormous progress in matter wave interferometry. The Colella-Overhauser-Werner (COW) type of neutron interferometer and the Kasevich-Chu (K-C) atom interferometer, are the prototype of such devices and the issue of whether they are sensitive to relativistic effects has recently aroused much controversy. We examine the question of to what extent the gravitational red shift, and the related twin paradox effect can be seen in both of these atom and neutron interferometers. We point out an asymmetry between the two types of devices. Because of this, the non-vanishing, non-relativistic residue of both effects can be seen in the neutron interferometer, while in the K-C interferometer the effects cancel out, leaving no residue, although they could be present in other types of atom interferometers. Also, the necessary shifting of the laser frequency (chirping) in the atom interferometer effectively changes the laboratory into a free-fall system, which could be exploited for other experiments.

[1] Email address: greenbgr@sci.ccny.cuny.edu. (Direct inquiries to this address).
[2] Email address: Wolfgang.Schleich@uni-ulm.de.
[3] Email address: rasel@iqo.uni-hannover.de.

## 1. Introduction

In the early days of quantum theory, even before the advent of modern quantum mechanics, one of the chief techniques for exploring many of the concepts involved was the invention of interesting "gedanken" experiments[1], conceptual thought experiments that were considered to be too intricate to ever be performed in the laboratory. The invention of the neutron interferometer[2,3] suddenly allowed one to actually perform many of these gedanken experiments and to verify experimentally some of the most unexpected, even bizarre, features of quantum mechanics. One of these experiments was the extremely beautiful and seminal COW experiment[4], which was carried out over 35 years ago and which for the first time saw a matter wave interference effect in the laboratory due to gravity. (We shall refer to the Colella-Overhauser-Werner neutron interferometer as the COW interferometer, and the Kasevich-Chu fountain effect interferometer as the K-C interferometer.)

Since then, more advanced matter wave interferometers[5-9] have become feasible, and the elegant atomic fountain experiments of Chu and his collaborators[10,11] have achieved such an accuracy and versatility that they are sensitive to many details of free-fall and equivalence principle effects that the old neutron interferometers could barely see. Actually, this is just the vanguard of a new field exploring experimentally the interface between quantum mechanics and gravity. Since there is no general theory connecting the two fields such experiments will become more and more important in exploring this new terrain[12]. Newer developments include such phenomena as the free-fall experiments on Bose-Einstein condensates and other experiments that can be done in space[13], bound states of neutrons in the earth's gravitational fields[14,15], and gravitationally produced phase shifts[16]. The advent of satellites and rockets combined with atom optics will also introduce a new range of possible effects, such as the Lense-Thirring effect[17-19], that can be explored as the field expands.

Even when the COW experiment was first performed, the issue was raised as to whether one could see relativistic effects with a neutron interferometer, and was largely resolved[20,21]. Now, once again, this question has come up, and is being warmly debated, in the form of what relativistic effects can and have been seen in the atomic interferometers. Although the physical pictures are quite different, the conceptual problems that arise in both cases, the neutron and atom interferometers, are quite similar, and yet the relativistic effects show up quite differently.

We shall discuss the problem of which relativistic effects can be seen in either type of matter wave interferometer. Although the accuracy of the atom interferometer experiments makes them useful for testing the validity of the equivalence principle, we will assume for our purposes that relativity and the equivalence principle hold. Therefore we only need the equivalence principle in Einstein's original sense, that all bodies fall at the same rate in a gravitational field, regardless of their composition, and that locally, inertial forces are equivalent to gravitational forces, and that a frame in free fall acts as an inertial system. We will not need the more modern formulations, which are aimed at testing the theory (see Ref. (22)). In particular though, we will discuss the claim of Müller, *et al.*[23-25], that one can measure the gravitational red shift in atom-interferometer experiments with a much greater precision than has been seen in other measurements of the red shift. On the other hand, Wolf, *et al*[26-28], have claimed that it is impossible to see such a red shift in matter wave interferometer experiments. We will point out one can in fact see the red shift, and indeed one can also see a second and related relativistic effect, the twin paradox, in matter wave interferometers, but in the K-C atomic interferometer these effects cancel out. These two effects are relevant to both the neutron and atomic interferometer experiments, but even when they do appear, what one is actually measuring is the non-relativistic residue of each of these effects, which is non-vanishing but which does not need relativity for its explanation. (In many quantum mechanical effects, proper time shows up as a phase factor, and often in the non-relativistic limit these accumulated phase effects add up and can be seen experimentally in interference experiments. That is what is happening here. See Ref. (29).)

While a relativistic point of view provides an insight that is not available non-relativistically, these effects are nonetheless interpretable as the result of non-relativistic forces. And in fact the phase shifts can be derived without any relativistic considerations at all[30,37]. The relativistic point of view does not provide any extra precision and we shall actually show that in the atomic fountain experiments as performed, the relativistic effects cancel out, so that in this respect we think that the claims of Müller, *et al.,* are not correct. And in fact, by continuously shifting their laser frequencies (chirping), they were effectively converting the laboratory into a free-fall, inertial system, in which no gravitational force could be felt. On the other hand, this does not happen in the neutron experiments, which do see the non-relativistic residue of these relativistic effects. This represents a distinct difference between these two kinds of interferometers.

In order to reconcile the various methods of calculating the phase shift in the atomic interferometer, we have also performed a detailed representation-free calculation of these experiments, to be published elsewhere[37]. In this paper, we shall try to emphasize the physical meaning of the differences between the interferometers.

## 2. The Phase as Proper Time

As Müller, *et al*[23]., point out, and in fact as was already pointed out by De Broglie[38] in his thesis (1924), the phase φ of the wave function is well-known to be a measure of the proper time along the path of the particle. We can see this easily for a free particle. If we follow the phase along the center of the wave packet describing the particle,

$$\psi = \psi_0 e^{i(p\cdot r - Et)/\hbar}, \tag{1}$$

we can substitute

$$p = mv\gamma, \quad E = mc^2\gamma, \quad r = vt, \quad \gamma = (1 - v^2/c^2)^{-1/2}, \tag{2}$$

and get

$$\psi = \psi_0 e^{i(mv\gamma\cdot vt - mc^2\gamma t)/\hbar} = e^{imc^2(v^2/c^2-1)\gamma t/\hbar} = \psi_0 e^{-imc^2\tau/\hbar}, \tag{3}$$

where $\tau = t/\gamma$. More generally, if there is also an external gravitational potential present, the phase is

$$\begin{aligned}\varphi &= -\frac{mc^2}{\hbar}\int d\tau = -\frac{mc^2}{\hbar}\int\sqrt{g_{\mu\nu}dx^\mu dx^\nu} = -\frac{mc^2}{\hbar}\int\sqrt{\left(1+\frac{2U}{c^2}-\frac{v^2}{c^2}\right)}dt \\ &\xrightarrow[NR]{} -\frac{mc^2}{\hbar}t + \frac{1}{\hbar}\int(\tfrac{1}{2}mv^2 - mU) = -\frac{mc^2}{\hbar}t + \frac{1}{\hbar}\int L_{NR}\,dt,\end{aligned} \tag{4}$$

so that the first order correction is just the non-relativistic Lagrangian, $L_{NR}$, controlling the classical motion of the particle, and it is independent of c. Here $U$ is the gravitational potential. This comes about because for weak fields,

$$\begin{aligned}&g_{00} = 1 + \frac{2U}{c^2}, \quad g_{ii} = -1, \\ &\sqrt{g_{\mu\nu}dx^\mu dx^\nu} = \sqrt{g_{\mu\nu}\frac{dx^\mu}{dt}\frac{dx^\nu}{dt}}\,dt = \sqrt{\left(1+\frac{2U}{c^2}-\frac{v^2}{c^2}\right)}\,dt.\end{aligned} \tag{5}$$

The Lagrangian can be expressed in terms of the Hamiltonian as

$$\int L_{NR}dt = \int(p\cdot\dot{r} - H)dt = \int p\cdot dr - Et, \tag{6}$$

when energy is conserved.

Our proper time calculations will be based on this observation, that

$$d\tau = \sqrt{g_{\mu\nu}dx^{\mu}dx^{\nu}} \xrightarrow[NR]{} (1 + U/c^2 - \tfrac{1}{2}v^2/c^2)dt, \qquad (7)$$

where U is the external gravitational potential.

## 3. Gravitational Phase Difference in an Interferometer

We have just seen that the phase φ of the well-localized wave packet of a particle of rest mass *m* that is free or in an external gravitational field is a measure of the proper time, τ, along the path of the particle, such that $\varphi = -\frac{mc^2}{\hbar}\int d\tau$. In the non-relativistic limit

$$\varphi \xrightarrow[NR]{} -\frac{mc^2}{\hbar}t + \frac{1}{\hbar}\int L_{NR}dt, \quad \frac{1}{\hbar}\int L_{NR}dt \equiv -\frac{mc^2}{\hbar}\delta\tau, \qquad (8)$$

where τ = t + δτ, so that the first order contribution to the proper time is provided by the non-relativistic Lagrangian, $L_{NR}$, controlling the classical motion of the particle, and it is independent of c, the speed of light (from eq. (4)).

We remind the reader that even for a free particle, in order to identify the phase with the proper time, we had to substitute $r = v_{group}\, t$, so that the identification is accurate only at the center of the wave packet. In the general case we still need the trajectory of the particle, which is a classical concept that only makes approximate sense along the center of a well-defined wave packet, and the identification of the proper time is only valid in this restricted sense. It makes no sense for arbitrary points off the classical trajectory.

Eq. (8) represents the phase along each arm of the interferometer from the time the two parts separate until they meet again and interfere. In these experiments, what is measured is the difference between the phases along the two arms, $\delta\varphi_u - \delta\varphi_\ell$. The zero-order term, $mc^2t/\hbar$, which has a $c^2$ in it, acts as a sort of "zitterbewegung" term and drops out as it is common to both beams. The second term, due to the non-relativistic Lagrangian, causes the difference in proper times, $\delta\tau_u - \delta\tau_\ell$. In both the neutron and atomic interferometers, the two beams are at different heights for part of their trajectories, and this introduces a gravitational potential energy difference between the two beams, which by the red shift produces a difference in proper times between them. The two beams may also have slightly different velocities and this in turn causes a further difference in proper times due to the twin paradox effect.

There are many similarities between these two types of interferometers, despite the fact that the mechanism of the separation and recombination of the beams is quite different in both cases. Fig. (1) shows their essential features. In the neutron case (a), the neutrons are deflected by bouncing off the crystal planes at A, B, C, and are coherently recombined at D. (An excellent review article on the details of this process can be found in ref. (39).) The half-circles represent the detectors. In order to introduce gravity, the horizontal interferometer is rotated about the incident beam AC (although there are variations in different experiments). In this way the strength of the earth's effective gravity field is made to depend on the degree of rotation, and can be used as a variable parameter.

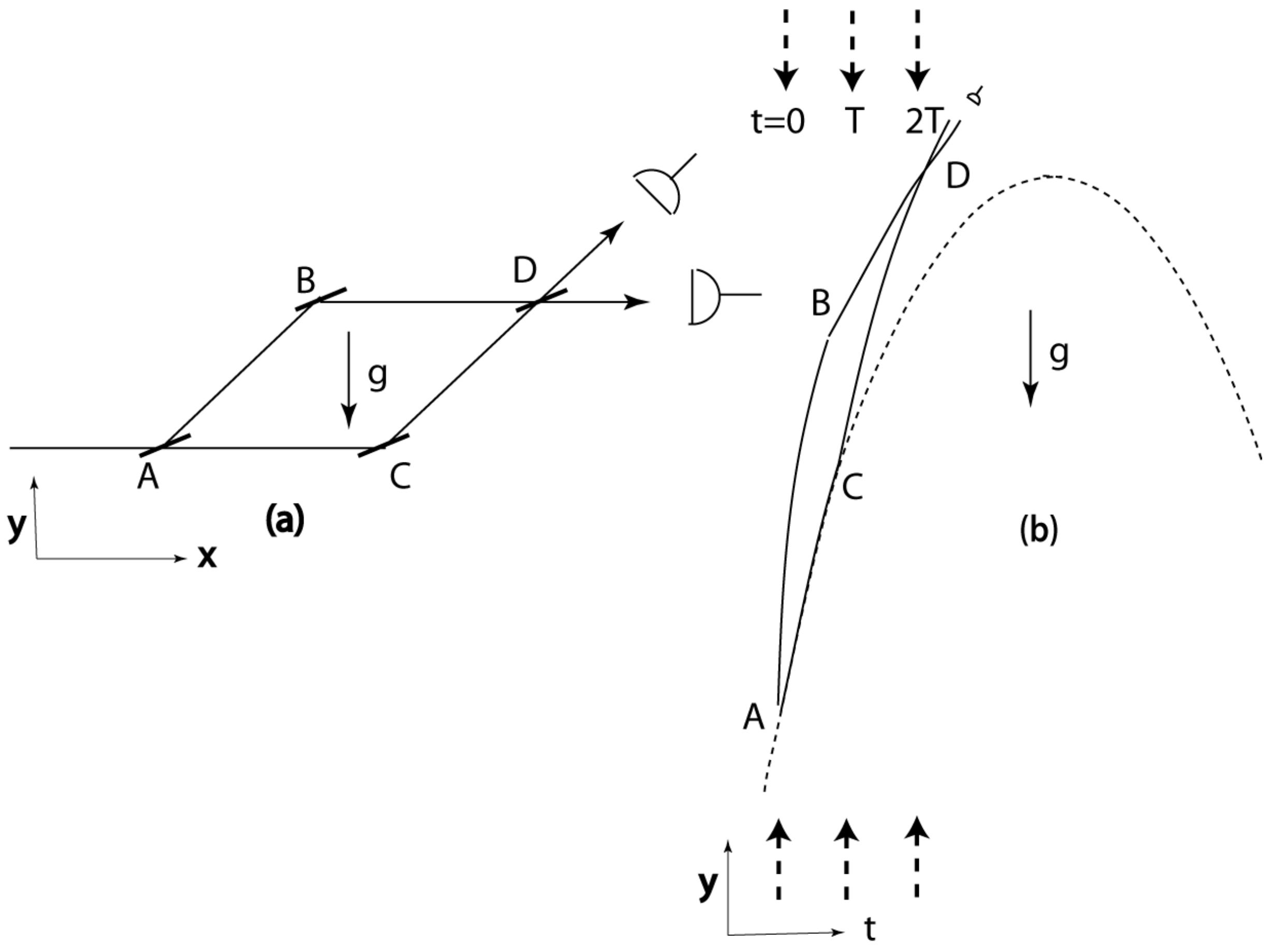

Fig. (1). Matter Wave Interferometers.
(a) The neutron interferometer: Coherent splitting of the beam is achieved by elastically scattering off the crystal planes at A, and the beams are then coherently recombined at D. Because the initially horizontal interferometer is rotated about the incident beam AC, both beams are at different heights, introducing a gravitational potential between them. (In this diagram we ignore the small bending due to gravity.)
(b) The atom interferometer: The beam is directed vertically upward. It is coherently split and recombined by absorbing a photon from a pulsed laser. Gravity is always acting. The dotted beam is the original path, and the solid beams are the interferometer beams created by the laser interactions. Here, time is plotted along the horizontal axis.

In the atomic interferometer case, Fig. (1b), the atoms are shot vertically upward against gravity and absorb photons from the lasers (the vertical dashed lines). They then recoil and are also recombined at D. The lasers are pulsed at times 0, T, and 2T. The pulse at time t = 0 separates the beams. The pulse at time T changes their directions, and the last pulse at time 2T coherently recombines them and determines how much of each beam enters each of the two detectors. The vertical momentum received by the atoms in absorbing the laser photons is analogous to the transverse momentum pulse received by the neutrons, in bouncing off the crystals. In Fig. (1b), for the atomic case, the horizontal axis represents time.

## 4. Differences Between the K-C Atom and COW Neutron Interferometers

In the K-C atom interferometer, momentum is conserved in the absorption or emission of the photon. When gravity is added, then each beam gets a gradual extra kick from gravity, each in the same direction, by $\delta k = m\delta v / \hbar, \delta v = gT$, and this momentum is not affected when the photon is absorbed (see Fig. (2)). So both beams fall like a free particle would, aside from the kick by the laser beam, and then when the beams recombine, the entire wave packet has fallen by the same amount a free particle would have, because the momenta due to the laser kicks cancel out, and both beams consume the same amount of proper time. (In fact, due to the laser kicks at time t = 0 the beams separate by distance $t\Delta v$, while the kicks at time T bring them together at the same rate. Gravity doesn't change this.)

The situation in the COW neutron interferometer is quite different. Because of this we are going to analyze a situation that does not represent the actual experiment for either the atom or neutron interferometer, which is depicted later in Fig. (4). It is instead a very similar but much more symmetric situation, one that represents an experiment that might have been carried out, but wasn't. It is more symmetric in that gravity points straight down in Fig. (4), and both halves of the diamond are symmetric. In the actual experiment (see Fig. (1)), the two halves of the diamond are tilted and the beams meet at a higher elevation than they started. The physical

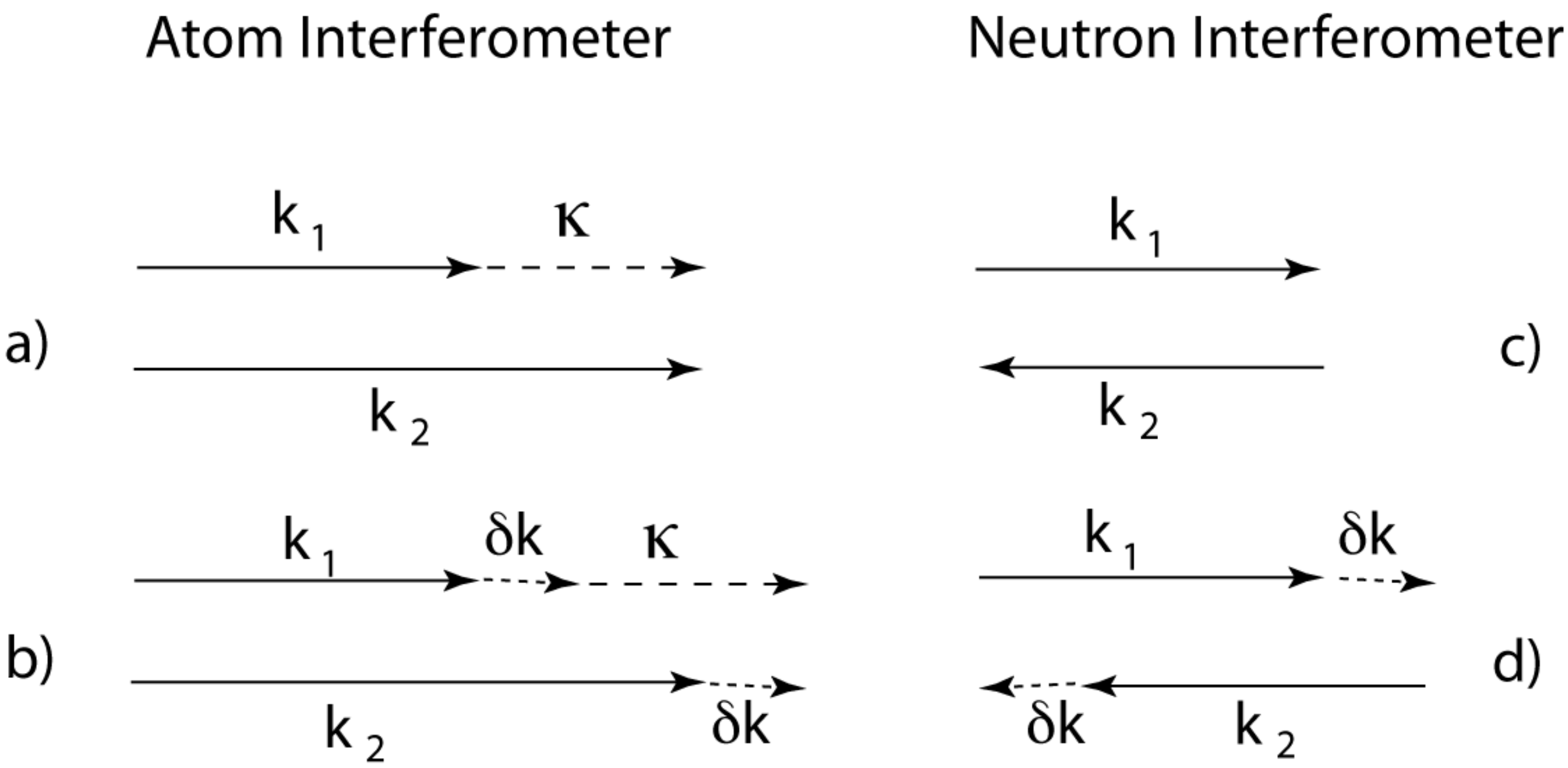

Fig. (2)

Fig. (2). Momentum Changes at Time T in the Matter Interferometer.

Atom Interferometer: a) No gravity- The Cesium atom ($k_1$) absorbs a photon (κ) and recoils ($k_2$). It thus does not conserve energy; b) Gravity present- The atom has some extra momentum due to falling, δk, which is passed on when it absorbs the photon.

Neutron Interferometer: c) No gravity- The neutron bounces elastically off the crystal atoms, thus conserving energy; d) Gravity present- The recoil off the crystal planes includes the momentum due to gravity, which is now in the opposite direction.

situations are very similar, but pedagogically it is much easier to analyze and to see what is happening in Fig. (4), and it brings out much more clearly the differences between the K-C and neutron interferometers, which we are trying to point out. The systems differ in that we have tilted the experiments and have made a constant velocity transformation to a system in which the two beams have vertical velocities $\pm v_{y0} = \frac{1}{2}\hbar\kappa/m$. This idealization will not affect the conclusions regarding the proper time and phase shift, that we shall draw in the two situations.

In the neutron interferometer there is no photon and the neutron is literally elastically bouncing off a line of real massive atoms, and because it has picked up a small extra momentum from gravity, this bounce includes and reverses the extra gravitational momentum (see Fig.(2)). So when the beams recombine they end up at the same height as they started (depicted in Fig. (7)), due to this elastic extra kick. If we look at this from the system freely falling with the beams, then it is the line of atoms in the crystal that is accelerating, and this kick is asymmetrical, depending on whether the atoms are effectively approaching or receding from the neutron. The neutron picks up a kick equivalent to twice this velocity of approach (or recession), and this does produce a difference in proper times elapsed between the upper and lower neutron beam, when they recombine.

Another way to say this is that in both cases, for the atom and the neutron, we are treating extreme situations, that of a very light particle bouncing off a very heavy one. The difference is that in the atom interferometer, it is the cesium atom that forms the beam, which is the heavy particle, while in the neutron interferometer it is the neutron, that forms the beam in this case, which is the light particle. The result is that the neutron bounces off the crystal and reverses all its momentum, including that of gravity. The situation is like that of bouncing a ball elastically off the floor. It reverses its momentum and rises to its original height. In the atom case, the photon flicks off the atom, or is absorbed by it, but it cannot reverse the motion of the heavy atom, which keeps falling under gravity, and so ends up falling by as much as if it had been dropped from rest, by $(1/2)gt_e^2$, where $t_e = 2T$ is the total elapsed time.

(We were amazed to have recently come across the following comment[40] in the early literature of the subject, after this paper was finished,: "The free evolution contribution...[in the atom interferometer]...turns out to be zero if there is no violation of the equivalence principle...this is in contrast to neutron interferometers...The distinguishing feature between the two cases is that the energy of the particle is not conserved during the light pulses of our [K-C] interferometer while it is conserved during the Bragg reflection processes of neutron interferometers." They do not explain this further, nor discuss the proper time. But this insight seems to have been forgotten over the intervening years by many practitioners in the field.)

There are also several further considerations that we should mention. When the neutron beam reaches the central ear of the interferometer it scatters off the planes of atoms perpendicular to the face of the ear (see Fig. (3)), called Laue scattering. This scattering takes place where the neutron hits the crystal at the time $T = L / v \cos \theta$, where

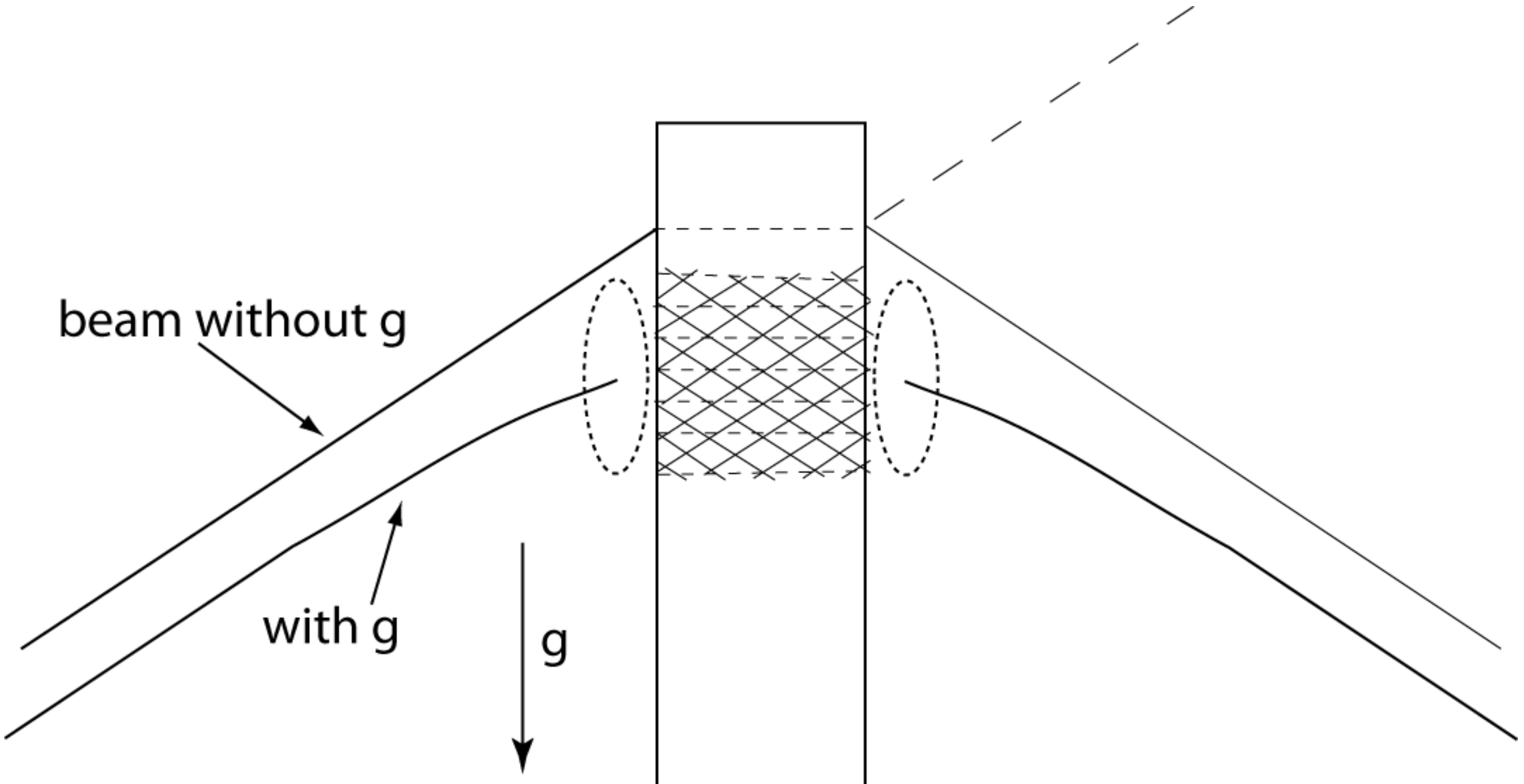

Fig. (3)

Fig. (3). Laue Scattering in the Crystal Ear. The neutron wave packet hits the crystal at time T, and scatters off the planes perpendicular to the face of the crystal, setting up standing waves inside the crystal, which convert back to traveling waves as it leaves the crystal. (This is different from hitting a single horizontal mirror at a fixed height, which would happen at different times, depending on the angle of the beam.) For simplicity we will ignore the thickness of the crystal in our analysis. See Refs. (3) and (21) for some more detail.

*L* is the distance between ears. Inside the crystal, the incident and reflected waves form standing waves between the crystal planes, and at the far end of the crystal they convert back into traveling waves. This wave picture is complementary to that leading to the conclusion above that the interaction looks like an elastic collision. (For simplicity we will ignore the width of the ear in our calculations.)

A further observation follows from the well-known fact that when one solves the Schroedinger equation in a periodic potential, one finds that there are energy bandgaps, regions where the neutron cannot propagate, and this is of course responsible for the Bragg scattering, at the regions between Brillouin zones. However this gap region has a finite bandwidth in energy, which was estimated in ref. (21) for a typical silicon neutron interferometer, to correspond to about $10^{-6}$ rad. in beam spread. If one assumes a velocity of about $10^5$ cm/sec for a thermal neutron leaving a nuclear reactor, and a spacing between the crystal ears of about 1 cm, the neutron will drop about $10^{-7}$ cm between ears due to gravity, which gives it a deflection of about $10^{-7}$ rad, well within the allowed spread. So the neutron will still be Bragg (or in this case Laue) scattered.

Another important effect in an atom interferometer is that the gravitationally falling atoms in the two beams that comprise the interferometer are slowly accelerating downward. This causes a Doppler shift that throws the atoms out of resonance with the two counter-acting laser beams. These two lasers set up a standing wave field that the atoms interact with. The atom is falling toward the lower laser and away from the upper one, and so the atom sees the lower laser as having too high a frequency, and the upper laser as having too low a frequency. In order for the lasers to stay in resonance with the falling atom one must raise the frequency of the upper laser and lower that of the lower laser (called "chirping" the frequency), in order to keep both of them in resonance with the falling atom. The effect of this is to essentially change the laboratory into the free-fall system as the lines of constant phase of the electromagnetic fields accelerate with the falling particles. (We discuss this at the end of the paper in Appendices G, H, and I.) But the atoms bouncing off these accelerating phase lines feel only a very small momentum kick, as it is a Doppler effect down by $\delta v \kappa / c$, where $\kappa$ is the wave number of the laser photons. and $\delta v \sim gT$. However there *is* an extra phase shift felt by the atoms due to this effect, but this does not contribute to the passage of proper time of the atoms, to first order.

## 5. Outline of Our Analysis

We are going to discuss the following four situations. First we will address the case of the atom interferometer at rest in the laboratory in a gravitational field. Then we will discuss the same situation as seen from a frame freely falling with the atoms. In this frame one does not see the gravitational field, according to the equivalence principle. However it is not a total free fall situation, as the lasers and the detectors are at rest in the laboratory, and so in the free fall system, they are accelerating upwards, and we will include the effect this has on the experiment.

Then we will discuss the situation for a neutron interferometer, first in the laboratory frame, and then in the free fall frame as before. In all these situations, we will be interested primarily in two questions: first, what is the difference in the passage of proper time between the two arms of the interferometer, and second, what will be the phase shift measured by the detectors.

We shall then interpret our results in terms of both the gravitational red shift and the twin paradox, and how these experiments relate to them. In these analyses, there are several effects that enter and that have to be taken into account. First, in the laboratory system, the beams are falling, and so sweeping past the lines of constant phase of the lasers, or the atomic planes of the neutron interferometer. On top of this, the laser frequencies are being chirped, which must be taken into account. At first we will ignore the chirping, and analyze the situation as though the frequencies were constant, and then show how to compensate for this effect. This will change the interpretation of the atomic interferometer experiment. Another complicating effect in the atom interferometer is that the atom beam does not pick up a momentum $\hbar\kappa$ by simply absorbing a photon, but rather picks up the momentum in a two-stage process, a Raman transition. The Raman transition is discussed in the appendices G, H, and I. This feature does not affect the validity of our results, or of our discussion of the Doppler effect.

A further phenomenon enters in the free fall frame. In this frame, the effect of gravity does not enter, but in the case of the atomic interferometer, as we have mentioned, the lines of constant phase of the effective laser crystal lattice are accelerating upward. Similarly, in the case of the neutron interferometer the real atomic planes in the interferometer are accelerating upwards. This has a very weak effect on the momentum transfer in the atomic interferometer (down by $\delta v/c$), but a very important one in the neutron interferometer, and comprises a major difference between the two cases.

Another effect also enters. In the *neutron* interferometer, the two beams meet at the third ear which then acts as a beam analyzer. It is sensitive to the relative phases of the two beams and determines what percentage of the beam goes to each of the two detectors. The analyzer is at rest with respect to the detectors and feeds the detectors. The detectors themselves, of course, are not sensitive to the detailed phase of the wave function, but they do not have to be. That role is played for them by the analyzer. In the *atom* interferometer, the absorption of the photon from the third laser pulse has the same effect, acting as an analyzer, determining the relative phase of the two beams and thus the relative intensity of the beams that travel toward each detector. The actual counting of the atoms or neutrons is made at the detectors, but the phase determination is made at the analyzer, which is at rest with respect to the detectors. In the laboratory system, where the lasers are at rest, the analyzer and detectors are also at rest. But in the free-fall system, which is determined by the falling beams, the analyzer point where the final phase is determined is not at rest. This point is accelerating upward, as are the detectors, and as it does so it is sweeping out a phase change that must be taken into account. We shall call this the *sliding effect*.

A final point should be made that offers an important check on our calculations. The phase shift where the beams recombine at the analyzer determines the number of particles that reach each of the two detectors, and so it is a physical invariant, and must be the same in both the laboratory and the free-fall system. In the free-fall system, as the reference frame falls with the atoms, there will be no gravity phase shift. But then one has the responsibility of finding out exactly what does cause the total phase shift. Clearly gravity is not the only effect. In some of our analyses, the crystal (or virtual optical crystal) is accelerating upward, and there will be both an acceleration and a sliding effect, and these in the free fall system will replace the effect that gravity has in the laboratory system. The proper time lapsed in each separate arm of the interferometer is also an invariant, and this too must be the same in both the laboratory and free fall systems.

Our calculations will show that both the red shift and the twin paradox effects show up in matter wave interferometers. But in the K-C interferometer they cancel out, and the two beams of the interferometer take the same proper time, while in the neutron interferometer they do not. So one can certainly see these effects in the neutron interferometer, and probably in differently designed atom interferometers that reflect the atomic beam.

## 6. The Proper Time and Phases in the K-C Atom Interferometer

### A. The Laboratory System

#### a. The Trajectories

We shall discuss the proper time here for the atom interferometer case, where the particle absorbs or emits a photon of momentum $\hbar\kappa$ , in the lab. system. We are working to lowest order in g. In the laboratory system, the trajectories of the particle along the different paths are shown in Fig. (4). The dashed lines are the trajectories the particle would have taken if there were no gravity acting. The laser pulses are at times 0, T, in the middle, and 2T, at the right hand side. Because of gravity, both bounces take place at a lower height than they would without gravity. The change in vertical momentum of each of the beams, on emitting or absorbing a photon, is $\hbar\kappa$ . This means the two beams at the left each have $v_{y0} = \pm\hbar\kappa / 2m$, and also that at the top and bottom and the far side, $\Delta k_y = 2k_y = \pm\kappa$ , for the atom interferometer in appendix A. (Note that in all the appendices we take c = 1.) From this we see that the center of the upper and lower beams recombine at 2T, and that the center of mass of the beam has fallen by $\delta y = -2gT^2 = -4d$, where $d = \frac{1}{2}gT^2$ . So the wave packet has fallen by the same amount as it would have if it were a free particle.

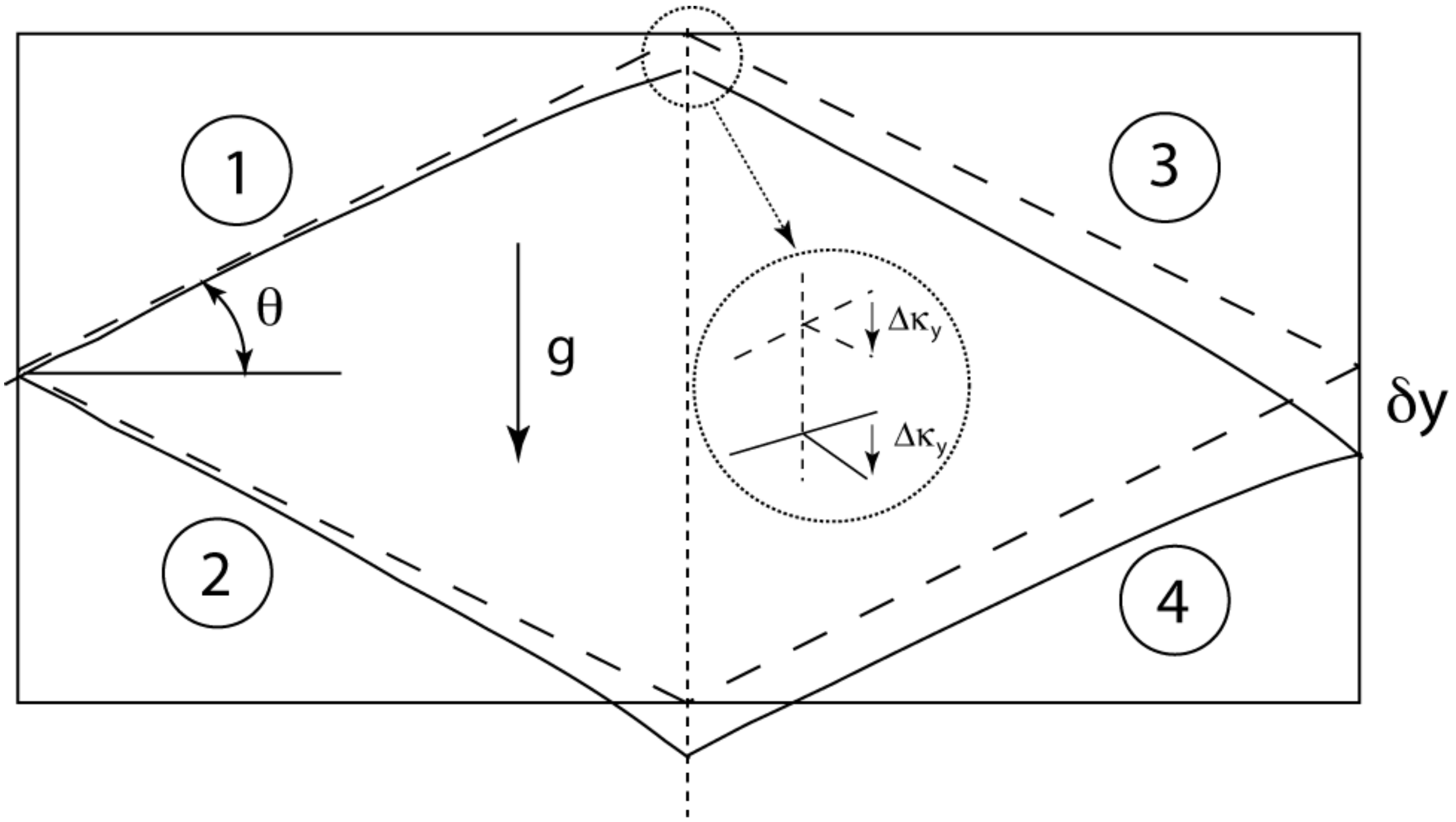

Fig. (4)

Fig. (4). The Trajectories of the Particles in the Atom Interferometer in the Laboratory. In the laboratory after the beams separate, with the upper beam absorbing a vertical photon of momentum $\hbar\kappa$ at time 0, they both fall under the influence of gravity. The upper beam re-emits the photon at time T, while the lower beam absorbs one, and they both keep falling. The lower beam re-emits it at time 2T where they recombine and continue on to one of the two detectors.

b. Proper Time in the Laboratory System

The general equation is, from eq. (7),

$$d\tau = [1 + U(y)/c^2 - v^2/2c^2]\, dt, \qquad (9)$$

in the non-relativistic limit. The details of the proper times taken by the two beams are worked out in appendix B, which indicates that in an atom interferometer, when the beams come together, there is no proper time difference between them (according to eq. (A11)), even though they have been at different heights. (This conclusion will be obvious in the free fall system.)

c. The Phases in the Laboratory System

However, when the two beams are recombined, they have both dropped by an amount $\delta y = -2gT^2$. (See eqs. ( A2) and (A4), and Fig. (4).) So the entire wave packet has fallen by this amount. But the detector has not fallen. It is fixed in the laboratory. So it is no longer sampling the center of the wave packet, but the phase has shifted by an amount equivalent to this displacement. One can calculate the equivalent phase shift from Fig. (5).

When we talk about a phase shift in scattering, we have to compare the different phases at the same point. This is what is measured by the detector. If you compare the phases at different points, there are many different factors that have to be considered, so it becomes much more complicated. One thing is that even without the perturbation, the phases at the two points are different and there are also time lags between the two points, all of which must be taken into account. Furthermore it should be noted that the phase shift at the detector is *not* the phase measured by the proper time, eqs. (4) and (8), which is only defined at the center of the wave packet (see the discussion below eq. (8)). In the K-C interferometer the proper time difference is zero, but the phase shift as measured by the detector is not zero, but is given below by eq. (12 ).

As an example, look at a wave impinging at a point, Figs. (5a,b). If it is retarded by x, where $\ell$ is the distance along the screen where the next wave will hit, then $\dfrac{x}{\lambda} = \dfrac{\delta\varphi}{2\pi}$. The distance along $\ell$ will be given by

$$\lambda = \ell \sin\theta, \quad \frac{x}{\delta\ell} = \sin\theta = \frac{\lambda}{\ell}, \quad \frac{x}{\lambda} = \frac{\delta\ell}{\ell}. \tag{10}$$

So the distance $\ell$ corresponds to a shift of one wavelength. And a shift by $\delta\varphi$ corresponds to a sliding by $\delta\ell$. Now look at two waves impinging on a point, at angles $\theta$ and $-\theta$, Fig. (5c). If both beams are retarded by the same amount they will still be in phase at the same point. But if one beam loses $\delta\varphi$ and the other beam gains $\delta\varphi$, a look at the diagram shows that the point where there will now be a maximum has shifted by $\delta\ell$. This corresponds to a shift by $\delta\varphi$ in the probability distribution, even though the relative phase difference between the beams is $2\delta\varphi$. You can see this analytically as

$$\psi = (e^{ik\cdot r} + e^{ik'\cdot r}) = 2e^{ik_x x}\cos k_y y, \quad |\psi|^2 = 4\cos^2 k_y y;$$
$$\psi' = (e^{i(k\cdot r+\alpha_1)} + e^{ik'\cdot r+\alpha_2)}) = 2e^{ik_x x}e^{i(\alpha_1+\alpha_2)/2}\cos[k_y y + (\alpha_1-\alpha_2)/2], \qquad (11)$$
$$|\psi'|^2 = 4\cos^2[k_y y + (\alpha_1-\alpha_2)/2].$$

This says that if you alter each of the phases, so the phase difference will be $(\alpha_1 - \alpha_2)$, the difference in the probability distribution will only be $(\alpha_1 - \alpha_2)/2$.

Here we can see that when the vertical distance $\delta y$ from where the two beams meet is changed by $\ell$, where $\ell \sin\theta = \lambda$, the phase shift $\varphi$ changes by $2\pi$. So,

$$\begin{aligned} k_y &= k\sin\theta = \frac{2\pi}{\lambda}\sin\theta = \frac{2\pi}{\ell}, \\ \kappa &= 2k_y = \frac{4\pi}{\ell}; \\ \frac{\varphi}{2\pi} &= \frac{\delta y}{\ell} = \frac{\kappa\,\delta y}{4\pi}, \\ \varphi &= \frac{\kappa\,\delta y}{2} = -\kappa g T^2 = -R. \end{aligned} \qquad (12)$$

Note that $2k_y = \kappa$, as can be seen in Fig. (4), and $\delta y = -2gT^2$, which comes from eq. (A2) and (A4). We emphasize that this phase shift is precisely due to the beams having fallen by the classical amount, $y = -\frac{1}{2}gT^2$.

So this phase shift comes from the fact that the wave packet has shifted. Note that this shift $\delta y$ is linear in $g$, and so the change in potential $\delta U = g\ \delta y$, due to it, is of second order in $g$, and so this does not change the calculation of the proper time.

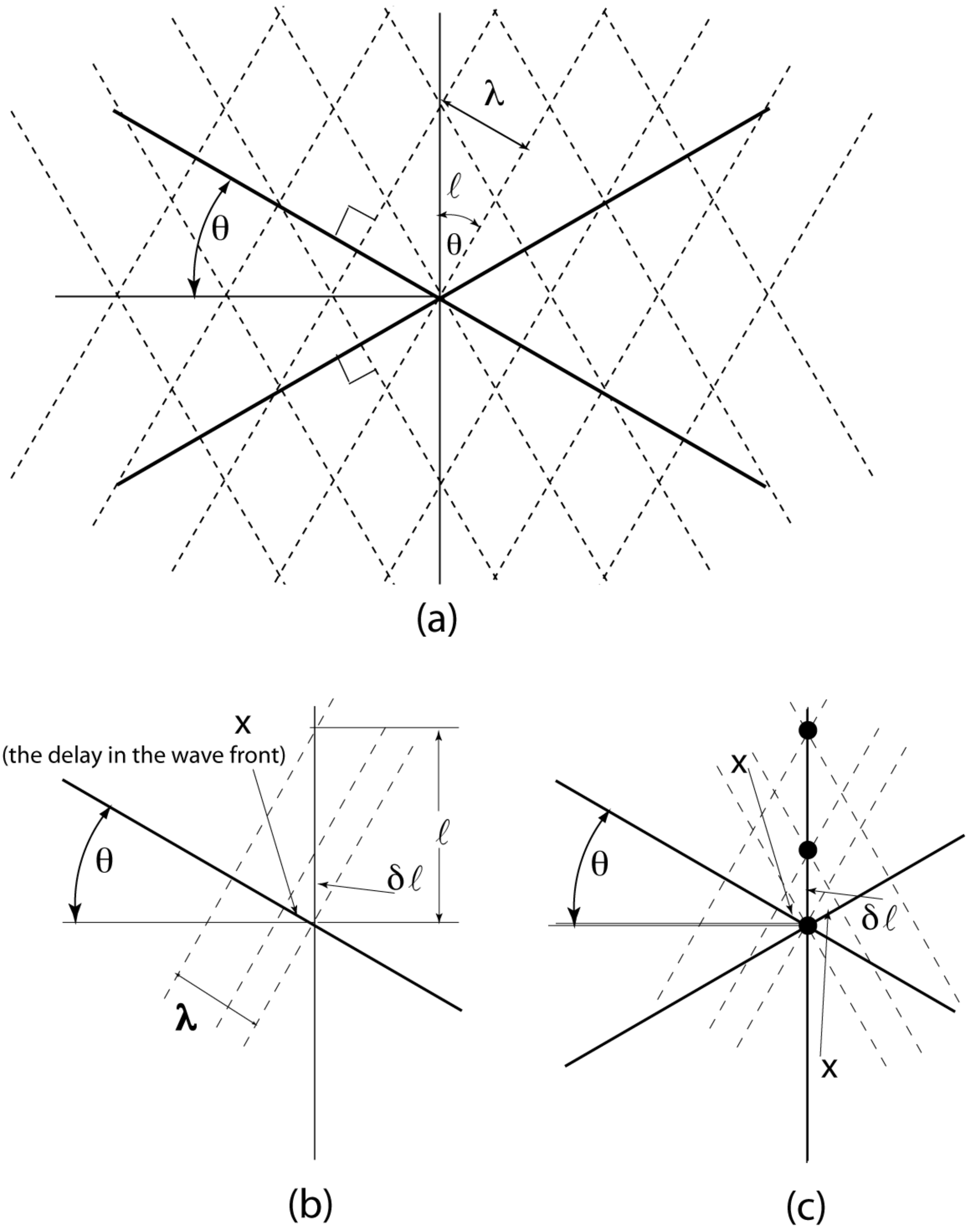

Fig. (5)

Fig. (5). Converting distance to phase. (a) The wave vectors of the two incoming atomic beams each have an angle θ with respect to the horizontal line, and are indicated by solid lines. The corresponding wave fronts, depicted by dashed lines, are orthogonal to these wave vectors. From where the two beams meet, the vertical distance is given by δy. Every time δy changes by $\ell$, where $\ell \sin\theta = \lambda$. the phase φ changes by 2π.

(b) When one beam is advanced by the distance *x*, where $x/\lambda = \delta\varphi/2\pi = \delta\ell/\ell$, its phase is advanced by $\delta\varphi$.

(c) If one beam is advanced by x, while the other is retarded by x, the maximum will still be shifted by $\delta\varphi/2\pi = x/\lambda = \delta\ell/\ell$ along the screen, even though the two beams differ in phase from each other by 2δφ. (See eq. (11)).

## B. The Free Fall System in the K-C Interferometer

### a. Trajectories and Proper Time in the Free Fall System

In the laboratory system, gravity is acting and distorts the trajectory of the beam. In the free fall system, the whole system is accelerating downward with an acceleration *g*, and one does not feel gravity acting. So the trajectories in this system are straight lines. The only way in which this system differs from a true inertial system is that the lasers are at rest in the laboratory, and so are accelerating upward in this system. In the laboratory the counter-propagating laser fields produce standing waves which are at rest in the laboratory. The difference is that the lines of constant phase are accelerating upward in the free fall system of the beams. These accelerating fields do not make any difference to lowest order, on the momentum of the photons that are absorbed, as they are Doppler-shifted by $\kappa \to \kappa(1+v/c)$ to lowest order, which is way down, compared to a particle rebounding elastically off a solid wall (as in the neutron interferometer) by a factor $\dfrac{\delta p_{atom\,int.}/p_0}{\delta p_{neutron\,int.}/p_0} = \dfrac{\delta v/c}{m\delta v/mv_0} = \dfrac{v_0}{c} \approx 10^{-5}$, where $\delta v \sim gT$.

So in the free fall system, there are no proper time effects due to the gravitational field, or to velocity differences, as shown in Fig. (6). This result agrees with eq. (A11), since the difference in proper time is an invariant. Fig. (6) also explains why the particles fall like a free particle in the laboratory, because nothing is happening to the beams in the free fall frame, while the laboratory is accelerating upward.

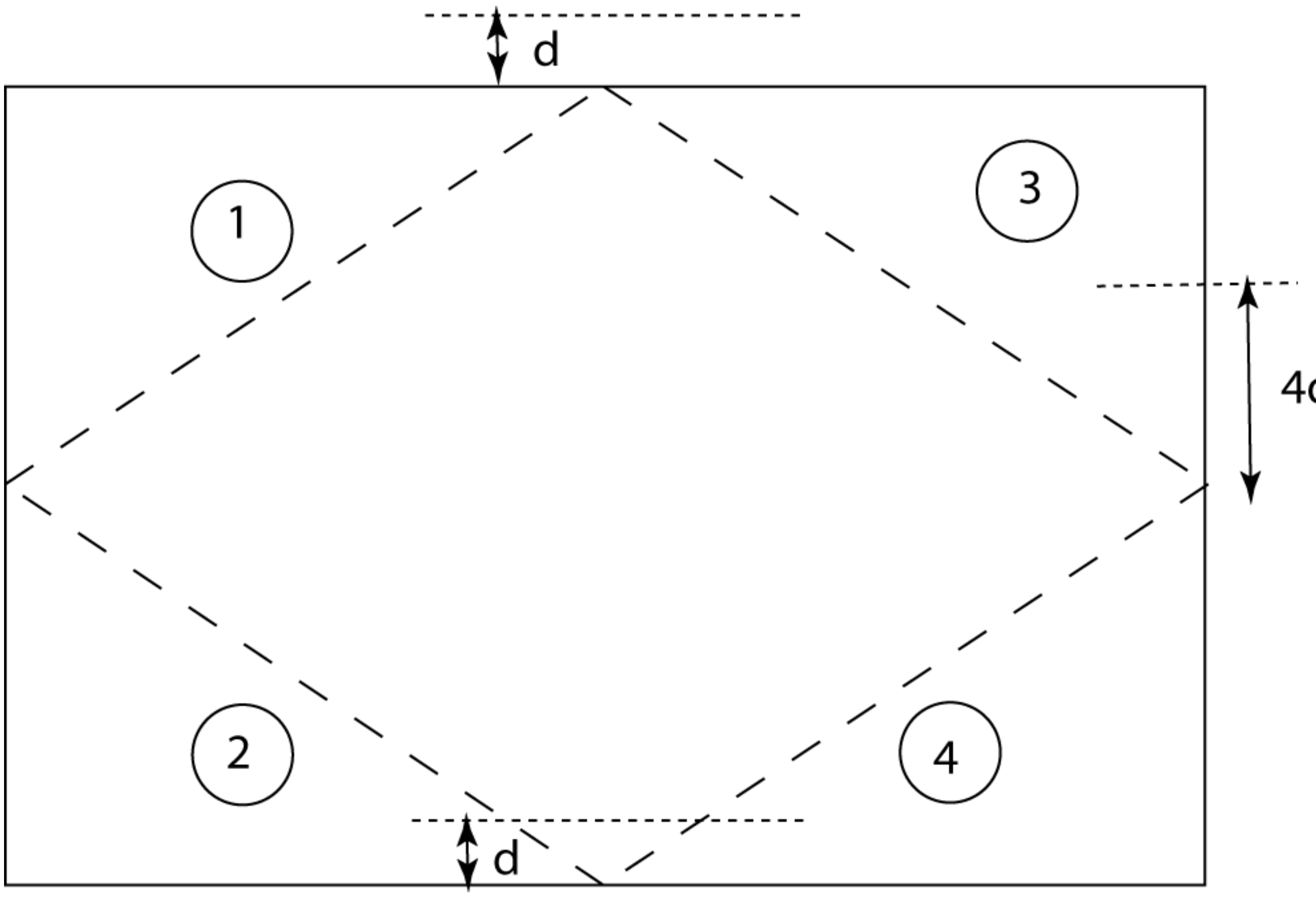

Fig. (6)

Fig. (6). The Atom Interferometer in the Free Fall System. There is no gravity in this system, and the atoms in each beam move as though it was an inertial system. But the lasers and detectors, which are at rest in the laboratory, are accelerating upward in this system, as are the lines of constant phase of the lasers, which are shown here by the closely dashed lines. They are shown displaced by the distance d = $gT^2/2$, that they have moved by time T. After time 2T, they have moved by 4d.

b. The Phases in the Free Fall System

The only difference between this case and a true inertial system is that the lasers and detectors are also at rest in the laboratory, and the lines of constant phase are accelerating upward in the free fall system, as shown by the closely dashed lines in Fig. (6). In the atom interferometer, after time T, at the top and bottom of Fig. (6), even though the lines of constant phase of the E-field of the laser are accelerating upward, there is no Doppler effect due to the gravity force (i. e., it is smaller by $\delta v/c$), and the only effect is that at time 2T, the analyzer point and detectors have slid upward by $4d = \frac{1}{2}g(2T)^2$. Since the analyzer sliding upward is equivalent to the pattern sliding downward, this gives $\varphi = -\kappa g T^2$ as in the laboratory frame.

Another way to say this is that from the top lines of eqs. (A1) and (A3), $y_1 - y_2 = 2v_{y0}t$ , and from the top lines of eqs. (A2) and (A4), $y_3 - y_4 = 2v_{y0}(T-t)$, where t runs from 0 to T, so that in the laboratory the differences between the top and bottom beam are just the diamond that is the dashed lines in Fig. (6), so that in the free fall system the beams follow the straight lines of this diamond. Thus, the total phase lost is equivalent to dropping by the distance $4d = \frac{1}{2}g(2T)^2 = 2gT^2$ . This comes to $\varphi = -\frac{1}{2}\kappa(2gT^2) = -\kappa gT^2 = -R$ , from the bottom line of eq. (12).

We have been talking about the free fall system, assuming a constant frequency in the lasers. But in the actual experiment[8,10,11], this was not the case. In their case, the frequencies were "chirped", or continually changed. We show in appendix I that this is equivalent to the acceleration of the lines of constant phase in the crossed laser beams. Thus the chirping changes the laboratory system into a true free fall system. So in this experiment, the laboratory system has become an inertial system, and that is the key aspect of this experiment, and this is a property that ought to be exploited in further experiments. (See the discussion section.) In this case the lines of constant phase are sliding across the detector and thus φ is not changing.

## 7. Proper Time and Phases in the COW Neutron Interferometer

### A. The Laboratory System

#### a. Trajectories

The situation for the neutron interferometer is completely different. In the atom interferometer, when the atom absorbs or emits a photon, it absorbs or emits the momentum of the photon. But if the lines of constant phase of the photon are moving with some velocity *v* or accelerating, the Doppler effect on the photon is down by *v/c*, so there is effectively no extra recoil by the atom. However, in the neutron case, the neutron essentially bounces off a brick wall. If the wall is moving, there is a significant Doppler effect and the neutron picks up twice the speed of the wall.

Even if the wall is not perfectly reflecting, the reflected neutron will pick up twice the speed of the wall. One can see this by making a boost to the speed of the wall. In the boosted frame the wall is at rest, and one can fit the appropriate boundary conditions. To get to this frame one can make a Galilean transformation on the incident and reflected beams, $\varphi(x',t) = e^{i\frac{m}{\hbar}(v\cdot x' + v^2 t/2)}\psi(x,t), \quad x' = x - vt$ . Then the incident beam, $k$, becomes $e^{im[v\cdot x' + v^2 t/2]/\hbar} e^{i[k\cdot(x'+vt) - \hbar k^2 t/2m]} = e^{i[k'\cdot x' - \hbar(k')^2/2m]}$ , $k' = k - mv/\hbar$ . So in this frame, moving away from the beam, the beam momentum is given by $k'$. For the reflected beam, the mirror is static in the moving frame, so the reflected beam will be $k_r' = -k'$ . This will be equivalent to the reflected beam in the original system being given by $k_r = k_r' + mv/\hbar = -k' + mv/\hbar$ , or $k_r = -(k - 2\,mv/\hbar)$, so the reflected beam loses the velocity $2v$. This derivation is independent of the nature of the mechanics of the beam splitter used, or its material composition.

The trajectories for the upper and lower beams in the interferometer are worked out in Appendix C. The situation in the neutron interferometer starts out exactly as in Fig. (6). But after that, the system evolves as in Fig. (7).

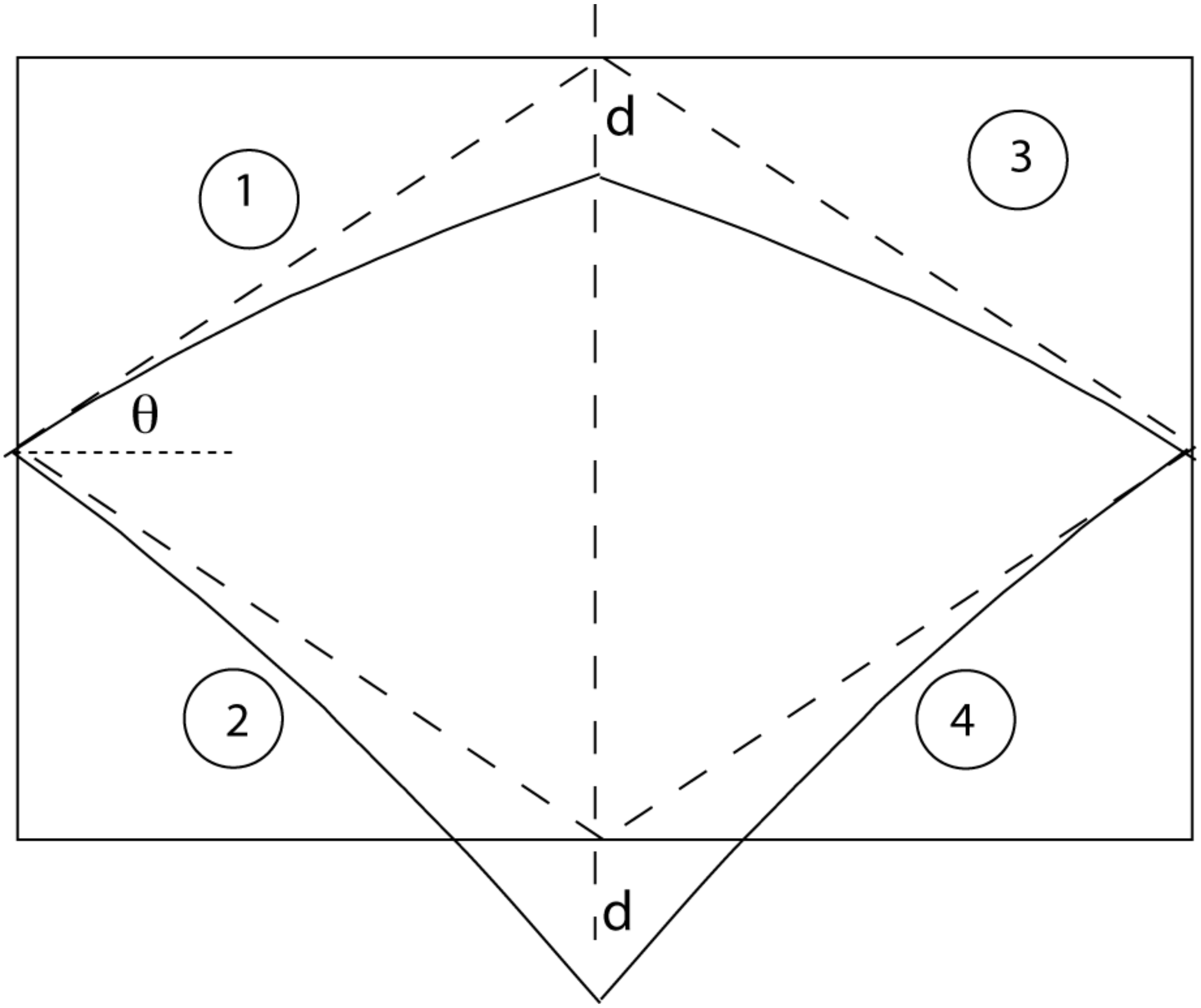

Fig. (7)

Fig. (7). The Neutron Interferometer in the Laboratory System. Without gravity acting, the beams would follow the dashed lines. With gravity, they both deflect downward. But in the elastic bounce at the center slab, the velocities, including that due to gravity, are reversed. So the beams return to their initial height at the final slab.

The beams are moving in the lab and being deflected downward by gravity. They have the same x-velocity, and so hit the central slab of the interferometer at the same time, but they are both falling downward in the y-direction, and so there is an asymmetry between the upper and lower beam. But at the central slab, the velocities are reversed, including the extra velocity picked up in falling gravitationally, which gives them enough velocity to return to their initial heights. This is so for both the upper and lower beam, which leads to a mirror symmetry about the middle ear, as can be seen from Fig. (7), and so there is no displacement of the beams at time 2T.

b. Proper Time in the Lab System

While there is no displacement of the two beams, there is a definite asymmetry in the proper times taken by each of them, which will be easiest to see in the free fall system (see Fig. (8)). The details for the proper times taken by the two beams are worked out in appendix D, starting from $d\tau = (1 + U/c^2 - v^2/2c^2)dt$, (taking c = 1, there). From this we see that, unlike the case for the atom interferometer, in the neutron interferometer there is definitely a difference in proper times. In the laboratory frame, half of the effect comes from the red shift, while the other half comes from the difference in velocities, or the twin paradox.

c. The Phases in the Lab System

For the purpose of calculating the proper time consumed, one has to follow the trajectory of the particle, but there is a much easier way to calculate the phase shift for small perturbations, for which we recall an important result[20]. From eqs. (6) and (8), we have the phase change, φ, along the path of one of the beams of the interferometer.

$$\varphi = \frac{1}{\hbar}\int p \cdot dr = 2\pi \int \frac{dr}{\lambda}. \qquad (13)$$

We do not include the zeroth-order contribution from the $mc^2$ term, which we have noted cancels out in an interference experiment. This equation assumes that energy is conserved, and says that we can follow the wavefronts and count the number of wavelengths.

When a perturbation is added to the system, the phase is changed by

$$\delta\varphi = \frac{1}{\hbar}\left(\int \delta p \cdot dr + \int p \cdot \delta(dr)\right). \qquad (14)$$

The first term in eq. (14) is the change of momentum calculated along the unperturbed trajectory. The second term contains the effects of the changing trajectory of the particle. It turns out to give no contribution to the phase. We emphasize that what we want is the change in phase at a given point, so that it is important that all changes be referred back to this point. In such circumstances, the second term gives no contribution whatsoever, to lowest order. We will discuss why this is so in Appendix F.

This result, that there is no contribution to the phase change from a shifting trajectory, is quite general for any continuous small perturbation, and should probably be called the "Golden Rule" of small perturbations. So the complete change due to the perturbation is contained in the first term of eq. (14). (This completely agrees with the calculations of Ref. (31)).

Using conservation of energy, the change in $p$ due to the added perturbation $\delta V$, will be

$$\frac{(p+\delta p)^2}{2m}+(U+\delta U)=E, \quad \frac{p\cdot\delta p}{m}=v\cdot\delta p=-\delta U, \\ \int \delta p\cdot dr=-\int \delta U\, dt. \tag{15}$$

We reiterate that we calculate this along the unperturbed trajectory. In eq. (15), this calculation gives the phase shift along each separate beam of the interferometer. In the actual interferometer, this equation generalizes to

$$\Delta\varphi \equiv \delta\varphi_u-\delta\varphi_\ell=-\frac{1}{\hbar}\int(\delta U_u-\delta U_\ell)dt=-\frac{1}{\hbar}\int \Delta U\, dt, \\ \Delta U=(\delta U_u-\delta U_\ell) \tag{16}$$

where the subscripts $u$ and $\ell$ refer to the upper and lower beams respectively. We have assumed energy conservation, which holds in the neutron interferometer. In the atomic case, the atom absorbs a photon from the laser and later re-emits it. These energy effects do not affect the validity of eq. (16). (See Appendix H.)

In our analysis, when gravity is considered as a small perturbation on the system, the entire effect is contained in eq. (16). This is very valid here, since the introduced gravitational potential energy is about $10^{-7}$ times the kinetic energy. In the laboratory frame, the interferometer is at rest and the neutron is falling under the influence of gravity. In this case, there will be a phase shift due to the gravitational field, the "red shift", a relativistic effect that induces a phase shift due to the difference in proper time between the two beams, caused by their different gravitational potentials. What is observed is the non-relativistic residue of this effect.

In this frame eq. (16) tells us to calculate the phase shift due to the different gravitational potential acting on each of the beams, and the effect is $-\frac{1}{\hbar}\int \Delta U\, dt$. In this situation, there is a there is a coherent amplitude for the neutron to be in either the upper or lower beam. These beams are separating with relative velocity $\Delta v = \hbar k / m$ , and the difference between their heights, *Δy,* grows as $t\Delta v$ , while they are moving in the linear potential $\Delta U = mg\Delta y$. (See the dashed trajectories in Fig. (7).) The phase difference, $\Delta\varphi$, that accumulates while they are separating by $\Delta y = t\Delta v = t(\hbar k / m)$ , reads

$$\Delta\varphi = -\frac{1}{\hbar}\int_0^T \Delta U\, dt = -\frac{1}{\hbar}\int_0^T mg\Delta y\, dt = -\frac{1}{\hbar}\int_0^T mgt\frac{\hbar\kappa}{m}dt = -\frac{1}{2}\kappa g T^2. \qquad (17)$$

and the same amount accumulates while they are coming together again, so that

$$\varphi_{tot} = -\kappa g T^2 = -R. \qquad (18)$$

The interferometer, as well as the analyzer point and the detectors, are at rest and so there is no further phase shift.

## B. The Free-fall System for the Neutron Interferometer

In order to check the result for the laboratory system, we shall calculate it in the system falling with the neutron beam. In this system, there is no gravitational field present, but the interferometer is accelerating upwards. (See Fig. (8)). This leads to the neutron taking the same paths 1 and 2 in this system, that it would take if there were no gravitational field present, but at the center slab, the planes of the crystal are accelerating upward, and the neutron will bounce off these crystal planes with a Doppler shift equal to twice the velocity of the planes, or 2gT. The upper beam hits a plane accelerating away from it, while the lower beam hits a plane accelerating toward it. The trajectories and proper times in this system are worked out in appendix E.

This result agrees with the one obtained in the lab system, eqs. (A13), (A14) and (A17). In the lab system the two beams returned to their initial heights, but here, because the interferometer is accelerating upward, the two beams rise by an equivalent amount, *4d*, equal to the displacement of the interferometer in sliding upward. The experiment is conducted in the laboratory system, and so we must add the phase gained by the interferometer in sliding upward by 4d in this system. This comes to the same numerical correction as was made in the lab system and one can see from Fig. (7) that the upper beam has lost phase relative to the lower beam, so these two effects add in the same way as before to give the same correction, and the same result. (See eq. (12) and the discussion surrounding it.) Thus the two calculations, made from different reference frames, agree in all respects.

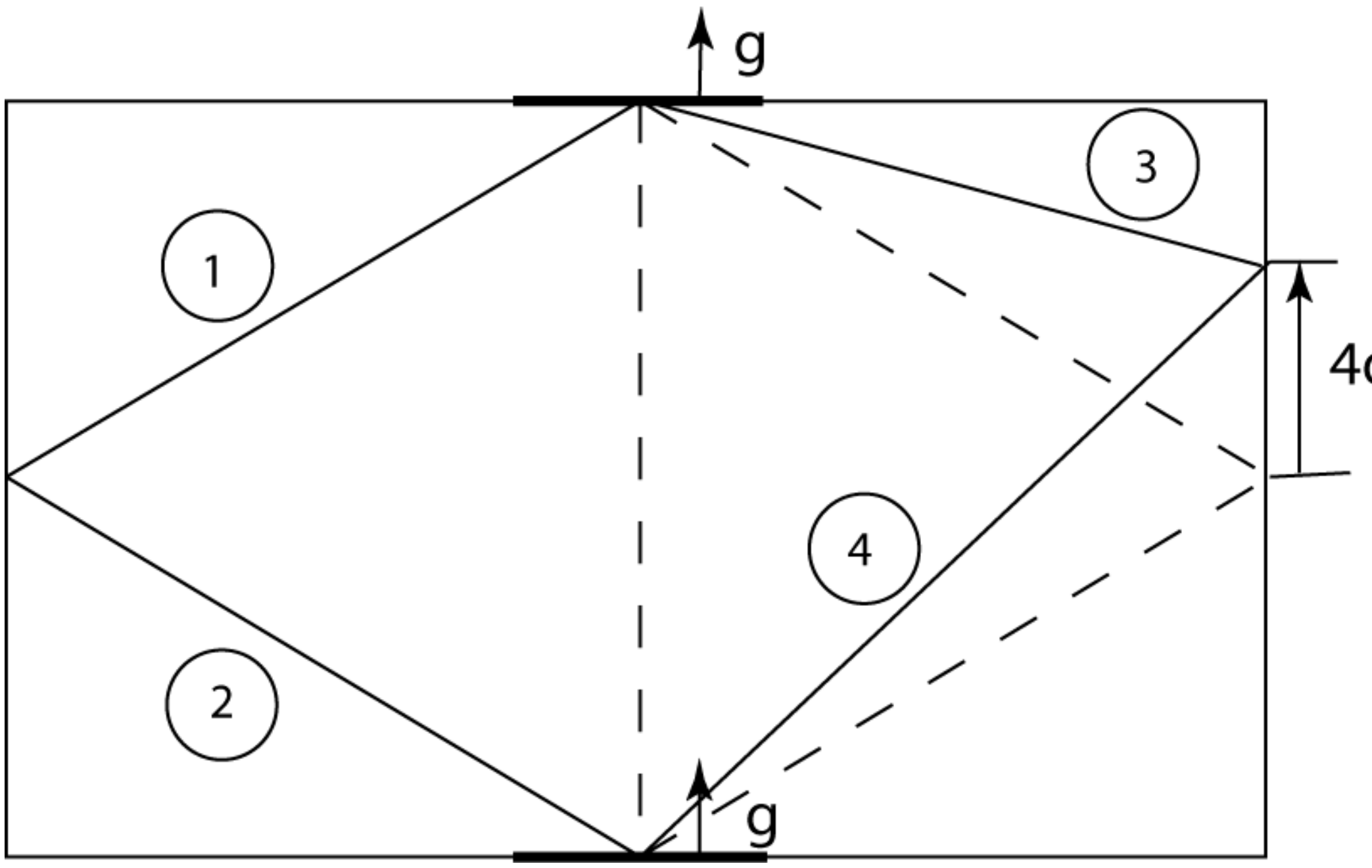

Fig. (8)

Fig. (8). The Neutron Interferometer in the Free Fall System. In the system freely falling with the neutrons, the neutrons travel in straight lines. However the crystal planes that they rebound from are accelerating upward, with acceleration $g$, and this causes an asymmetrical effect between the two beams, as the crystal is accelerating toward the lower beam, and away from the upper beam. The result is a difference in proper times taken by each of the two beams before they recombine.

## 8. Discussion

For the atom interferometer, so far as the trajectories are concerned, in the laboratory the particles fall as much as a free particle would (see Fig. (6)) and the phase arises from the sliding effect of the atoms with respect to the detector. In the free fall system conversely, it is the sliding effect due to the displacement of the detector, that would contribute to the overall phase if the frequencies of the lasers were kept constant. In the actual experiment[10,11], the chirping of the frequencies has the effect of accelerating the wave fronts so that they keep in pace with the falling beam (see appendix I), which eliminates this effect. In fact, the way they measured g was that they sat on a phase minimum as they shifted the laser frequency, or $R = \kappa g T^2 = \int \delta\omega \, dt = \frac{1}{2}\dot{\omega}T^2$, $\dot{\omega} = 2\kappa g$ .

So the same very ingenious technique of effectively eliminating gravity allowed them to measure g very accurately and also essentially converted the laboratory into a free fall system. This of course eliminated any relativistic effects they might otherwise have been able to see, but at the same time it created a situation that could be exploited in future experiments, namely the creation of a true inertial system in the laboratory. As a verification of this, an alternative derivation of the phase shift, using only Galilean invariance and non-relativistic concepts, is given in Ref. (30), which does not need to introduce the concept of the red shift, or of the Compton frequency, at all. A more rigorous version of this argument is given in our companion paper, Ref. (37).

In the case of the neutron interferometer, one can say that one is indeed seeing the red shift and the twin paradox, but one cannot prove that it is a relativistic effect, as it is only the non-relativistic residue of such an effect. In this case, relativity provides only an additional insight into what is happening, but it is not needed.

One other interesting lesson to be drawn from this is that even though the concept of proper time is not introduced in classical non-relativistic physics, in quantum theory proper time generally shows up as a phase and even in the non-relativistic limit, it leaves observable consequences in the form of accumulated phase shifts that we have referred to as the non-relativistic residue, which must be taken into account[29]. But once again, we would like to emphasize that the relativistic concepts, the red shift and the twin paradox, while they add interesting insights to help explain the physics involved in this problem, are not demanded in the analysis of it, as the results can equally as well be explained non-relativistically.

We will close the paper in Appendix J with some remarks concerning the various discussions of the red shift in the literature. Also we would like to note that our discussion of the atomic interferometers is specifically limited to the K-C interferometers. A number of people have pointed out to us that most of the experiments with neutron interferometers could also have been done with atom interferometers. For example, one could bounce a Cs atom off a wall of photons. But that would be a different experiment. To turn a Cs atom around in the laboratory with a single photon would require a gamma ray. Similarly, one could emulate the K-C atom interferometer by bouncing a phonon off the neutron. Again, this would be a different experiment.

Finally, we would like to bring up an interesting side issue.

In the Feynman path integral formulation one can, if one wants, break up the contribution into a kinetic energy part, a potential energy part, and a laser-interaction part, the first two of which cancel out, apparently leaving the entire effect to depend on the laser interaction[23,26]. Actually, it is the *difference* between the interactions of the two beams that enters here, in which gravity does play a role, which is why the phase shift depends upon $g$. This is clearly brought out in the representation-free operator calculation of our companion paper (see Ref. (37)).

## Acknowledgments

We would like to thank Achim Peters and Holger Müller for explaining some of the subtle points of the atomic fountain experiments to us, and to Endre Kajari and Anton Zeilinger for discussions. As part of the QUANTUS collaboration, this project was supported by the German Space Agency DLR with funds provided by the Federal Ministry of Economics and Technology (BMWi) under grant no. DLR 50 WM 0837. One of us (DMG) is grateful to the Alexander von Humboldt Stiftung for a Wiedereinladung which made this collaboration possible, and to the John Templeton Foundation, grant #21531, which made its completion possible.

# Appendices

## Appendix A: The Atom Interferometer-- Trajectories in the Laboratory

(In the appendices, we take c = 1, unless otherwise specified.)

In the different regions of Fig. (4), we have, starting from y = 0, where the beams split,

$$\begin{aligned} y_1 &= v_{y0}t - \tfrac{1}{2}gt^2, \quad v_1 = v_{yo} - gt, \\ y_1(T) &= v_{yo}T - \tfrac{1}{2}gT^2. \end{aligned} \tag{A1}$$

Also

$$\begin{aligned} y_3 &= (v_{yo}T - \tfrac{1}{2}gT^2) + (-v_{yo} - gT)t - \tfrac{1}{2}gt^2, \quad v_3 = (-v_{yo} - gT) - gt, \\ y_3(T) &= v_{yo}T - \tfrac{1}{2}gT^2 - v_{yo}T - gT^2 - \tfrac{1}{2}gT^2 = -2gT^2. \end{aligned} \tag{A2}$$

Here we have set the clock to 0 again, after t=T. The only change in the velocity of the beam after absorbing or emitting a photon, is the change at t=T, from $\pm v_{yo}$ to $\mp v_{yo}$. So $v_3(0) = v_1(T) - 2v_{y0}$ (remember we are resetting the clock for at time T, back to 0). It is also true that $v_4(0) = v_2(T) + 2v_{y0}$.

We also find

$$y_2 = -v_{yo}t - \tfrac{1}{2}gt^2, \quad v_2 = -v_{yo} - gt,$$
$$y_2(T) = -v_{yo}T - \tfrac{1}{2}gT^2, \quad v_2(T) = -v_{yo} - gT, \tag{A3}$$

and

$$y_4 = (-v_{y0}T - \tfrac{1}{2}gT^2) + (v_{yo} - gT)t - \tfrac{1}{2}gt^2, \quad v_4 = (v_{yo} - gT) - gt,$$
$$y_4(T) = -v_{yo}T - \tfrac{1}{2}gT^2 + (v_{yo} - gT)T - \tfrac{1}{2}gT^2 = -2gT^2. \tag{A4}$$

So both beams fall by the same amount, and land where a particle in free fall would have landed.

## Appendix B: The Proper Time in the Laboratory System

Using eq. (A1), we have for the beams in the different regions of Fig. (4),

$$d\tau_1 = [1 + g(v_{yo}t - \tfrac{1}{2}gt^2) - \tfrac{1}{2}(v_{yo} - gt)^2 - v_{x0}^2]dt$$
$$= [1 + gv_{yo}t - \tfrac{1}{2}v_{yo}^2 - \tfrac{1}{2}v_{xo}^2 + v_{yo}gt]dt, \tag{A5}$$

where we are working to first order in $g$. For $d\tau_2$ we have

$$d\tau_2 = [1 + g(-v_{yo}t - \tfrac{1}{2}gt^2) - \tfrac{1}{2}(-v_{yo} - gt)^2 - v_{x0}^2]dt$$
$$= [1 - gv_{yo}t - \tfrac{1}{2}v_{yo}^2 - \tfrac{1}{2}v_{xo}^2 - v_{yo}gt]dt, \tag{A6}$$

Overall, on the left side of the diagram, we have

$$\int (d\tau_1 - d\tau_2) = \int_0^T (2gv_{yo}t + 2v_{yo}gt)dt = 2gv_{yo}T^2. \tag{A7}$$

On the right side we have

$$d\tau_3 = \left\{1 + g\left[(v_{yo}t - \tfrac{1}{2}gt^2) + (-v_{yo} - gT)t - \tfrac{1}{2}gt^2\right] - \tfrac{1}{2}(-v_{yo} - gT - gt)^2 - \tfrac{1}{2}v_{x0}^2\right\}dt$$
$$= [1 + gv_{yo}T - gv_{yo}t - \tfrac{1}{2}v_{yo}^2 - \tfrac{1}{2}v_{xo}^2 - v_{y0}g(T + t)]dt. \tag{A8}$$

and

$$d\tau_4 = \left\{1 + g\left[(-v_{yo}t - \tfrac{1}{2}gt^2) + (v_{yo} - gT)t - \tfrac{1}{2}gt^2\right] - \tfrac{1}{2}(v_{yo} - gT - gt)^2 - \tfrac{1}{2}v_{x0}^2\right\}dt$$
$$= [1 - gv_{yo}T + gv_{yo}t - \tfrac{1}{2}v_{yo}^2 - \tfrac{1}{2}v_{xo}^2 + v_{y0}g(T + t)]dt. \tag{A9}$$

Thus

$$\int (d\tau_3 - d\tau_4) = \int_0^T [2v_{yo}gT - 2v_{yo}gt - 2v_{yo}g(T+t)]dt$$
$$= 2v_{yo}gT^2 - v_{yo}gT^2 - 2v_{yo}gT^2 - v_{yo}gT^2 \quad \text{(A10)}$$
$$= -2v_{yo}gT^2.$$

For the overall proper time difference, we then have

$$\Delta\tau_{upper} - \Delta\tau_{lower} = \int (d\tau_1 + d\tau_3) - \int (d\tau_2 + d\tau_4)$$
$$= \int (d\tau_1 - d\tau_2) + \int (d\tau_3 - d\tau_4) = 0. \quad \text{(A11)}$$

So there is no proper time difference between the two beams, calculated in the laboratory system.

## Appendix C: The Neutron Interferometer-- Trajectories in the Laboratory System

Referring to the various regions in Fig. (7), we find

$$v_1 = v_{yo} - gt, \quad y_1 = v_{yo}t - \tfrac{1}{2}gt^2,$$
$$v_1(T) = v_{yo} - gT, \quad y_1(T) = v_{yo}T - \tfrac{1}{2}gT^2;$$
$$v_2 = -v_{yo} - gt, \quad y_2 = -v_{yo}t - \tfrac{1}{2}gt^2, \quad \text{(A12)}$$
$$v_2(T) = -v_{yo} - gT, \quad y_2(T) = -v_{yo}T - \tfrac{1}{2}gT^2.$$

At the center, the beams are deflected by the horizontal rows of atoms they encounter and the beams after the central slab (assumed thin) are the mirror image of the beams before the slab. The beams reach the center slab at time T, at which time we start again at t = 0,

$$v_3(0) = -v_{yo} + gT. \quad v_3(t) = -v_{yo} + gT - gt,$$
$$y_3(t) = v_{yo}T - \tfrac{1}{2}gT^2 + (-v_{yo} + gT)t - \tfrac{1}{2}gt^2, \quad \text{(A13)}$$
$$y_3(T) = v_{yo}T - \tfrac{1}{2}gT^2 + (-v_{yo} + gT)T - \tfrac{1}{2}gT^2 = 0.$$

So the upper beam returns to its original height.

$$v_4(0) = v_{yo} + gT. \quad v_4(t) = v_{yo} + gT - gt,$$
$$y_4(t) = -v_{yo}T - \tfrac{1}{2}gT^2 + (v_{yo} + gT)t - \tfrac{1}{2}gt^2, \quad \text{(A14)}$$
$$y_4(T) = -v_{yo}T - \tfrac{1}{2}gT^2 + (v_{yo} + gT)T - \tfrac{1}{2}gT^2 = 0.$$

And the lower beam also returns to its original height.

Appendix D: The Proper time in the Laboratory System

Referring to Fig. (7) for the various regions of the beams, and using eq. (9), we find for the top beam,

$$d\tau_1 = \left[1 + g(v_{y0}t - \tfrac{1}{2}gt^2) - \tfrac{1}{2}(v_{y0} - gt)^2 - \tfrac{1}{2}v_{x0}^2\right]dt,$$

$$\Delta\tau_1 = T + \tfrac{1}{2}v_{y0}gT^2 - \tfrac{1}{2}v_0^2T + \tfrac{1}{2}v_{y0}gT^2; \qquad (A15)$$

$$d\tau_3 = \left[1 + g\left[(v_{y0}T - \tfrac{1}{2}gT^2) + (-v_{y0} + gT)t - \tfrac{1}{2}gt^2\right] - \tfrac{1}{2}(v_{y0} + gT - gt)^2 - \tfrac{1}{2}v_{x0}^2\right]dt$$

$$\Delta\tau_3 = T + v_{yo}gT^2 - \tfrac{1}{2}v_{yo}gT^2 - \tfrac{1}{2}v_0^2T + v_{y0}gT^2 - \tfrac{1}{2}v_{yo}gT^2;$$

The integrated periods, Δτ, are evaluated only to first order in g. Also, above, $v_0^2 = v_{y0}^2 + v_{x0}^2$.

For the bottom beam,

$$d\tau_2 = \left[1 + g(-v_{y0}t - \tfrac{1}{2}gt^2) - \tfrac{1}{2}(-v_{y0} - gt)^2 - v_{x0}^2\right]dt,$$

$$\Delta\tau_2 = T - \tfrac{1}{2}v_{y0}gT^2 - \tfrac{1}{2}v_0^2T - \tfrac{1}{2}v_{y0}gT^2; \qquad (A16)$$

$$d\tau_4 = \left[1 + g\left[-v_{y0}T - \tfrac{1}{2}gT^2 + (v_{y0} + gT)t - \tfrac{1}{2}gt^2\right] - \tfrac{1}{2}(v_{y0} + gT - gt)^2 - \tfrac{1}{2}v_{x0}^2\right]dt,$$

$$\Delta\tau_4 = T - v_{y0}gT^2 + \tfrac{1}{2}v_{y0}gT^2 - \tfrac{1}{2}v_0^2T - v_{y0}gT^2 + \tfrac{1}{2}v_{y0}gT^2.$$

The differences in proper times between the beams is then given by

$$\begin{aligned}
&\Delta\tau_{upper} - \Delta\tau_{lower} \\
&= (\Delta\tau_1 + \Delta\tau_3) - (\Delta\tau_2 + \Delta\tau_4) \\
&= (\Delta\tau_1 - \Delta\tau_2) + (\Delta\tau_3 - \Delta\tau_4) \\
&= 2v_{y0}gT^2 + 2v_{y0}gT^2 = 4v_{y0}gT^2.
\end{aligned} \qquad \text{(A17)}$$

So for the neutron interferometer, we see that the two beams take a different amount of proper time before they recombine.

Appendix E: The Free Fall System for the Neutron Interferometer

a. The Trajectories in the Free Fall System

For the trajectories in the free fall system, we refer to Fig. (8). Here there will be a definite asymmetry due to the accelerating mirrors. On rebounding from the crystal planes, as explained in the text, the beams will pick up twice the velocities of the mirror. For the upper beam,

$$
\begin{aligned}
&v_1 = v_{y0}, \quad y_1 = v_{y0}t, \\
&y_1(T) = v_{y0}T, \\
&v_3(0) = -v_{y0} + 2gT = v_3(t), \\
&y_3(t) = v_{y0}T + (-v_{y0} + 2gT)t, \\
&y_3(T) = v_{y0}T + (-v_{y0} + 2gT)T = 2gT^2 = 4d,
\end{aligned} \tag{A18}
$$

So the beam is deflected upward. For the lower beam,

$$
\begin{aligned}
&v_2(0) = -v_{y0}, \quad y_2(t) = -v_{y0}t, \quad y_2(T) = -v_{y0}T, \\
&v_4(0) = v_{y0} + 2gT = v_4(t), \\
&y_4(t) = -v_{y0}t + (v_{y0} + 2gT)t, \\
&y_4(T) = -v_{y0}T + (v_{y0} + 2gT)T = 2gT^2 = 4d.
\end{aligned} \tag{A19}
$$

This beam is deflected upward by the same amount as the upper beam, by the amount 4d, where $d=gT^2/2$, which is the amount by which the accelerating interferometer has risen in time T. This is consistent with eqs. (A13) and (A14), which say that in the lab system, which is accelerating upwards with respect to the free fall system, the beams have no overall deflection.

b. Proper Time in the Free Fall System

Again referring to Fig. (8), we can also calculate the proper times in the free fall system. Until time T, both the upper and lower beams take the same amount of proper time,

$$
\begin{aligned}
&d\tau_1 = [1 - \tfrac{1}{2}v_{y0}^2 - \tfrac{1}{2}v_{x0}^2]dt = (1 - \tfrac{1}{2}v_0^2)dt, \\
&d\tau_2 = (1 - \tfrac{1}{2}v_0^2)dt = d\tau_1, \\
&\Delta\tau_1 - \Delta\tau_2 = 0.
\end{aligned} \tag{A20}
$$

The second part of the trip is more complicated for both beams, because of the asymmetric recoils, and gives

$$
\begin{aligned}
&d\tau_3 = [1 - \tfrac{1}{2}(-v_{y0} + 2gT)^2 - \tfrac{1}{2}v_{x0}^2]dt, \\
&\Delta\tau_3 = (1 - \tfrac{1}{2}v_0^2)T + 2v_{y0}gT^2; \\
&d\tau_4 = [1 - \tfrac{1}{2}(v_{y0} + 2gT)^2 - \tfrac{1}{2}v_{x0}^2]dt, \\
&\Delta\tau_4 = (1 - \tfrac{1}{2}v_0^2)T - 2v_{y0}gT^2; \\
&\Delta\tau_{total} = \Delta\tau_3 - \Delta\tau_4 = 4v_{y0}gT^2.
\end{aligned} \tag{A21}
$$

This result agrees with the proper time difference between the beams that we calculated in the laboratory system.

Appendix F: The "Golden Rule" of Small Perturbations--No Phase Shift Is Produced by the Bending of the Beam

The shift in trajectory caused by a small perturbation (the second term in eq. (14)), does not contribute to the change in phase between the two beams. In order to see this, refer to Fig. (9), which shows that if there were no perturbation, the trajectory of the particle would have gone directly from point P to point A. With the perturbation present, the path gets deflected along the dashed line, D, to point B. (B is an arbitrary point along the deflected path.) The phase of the particle at B will be different from the phase at point A. The circular arcs in Fig. (9) represent the lines of constant phase of the unperturbed wave emanating out from point P.

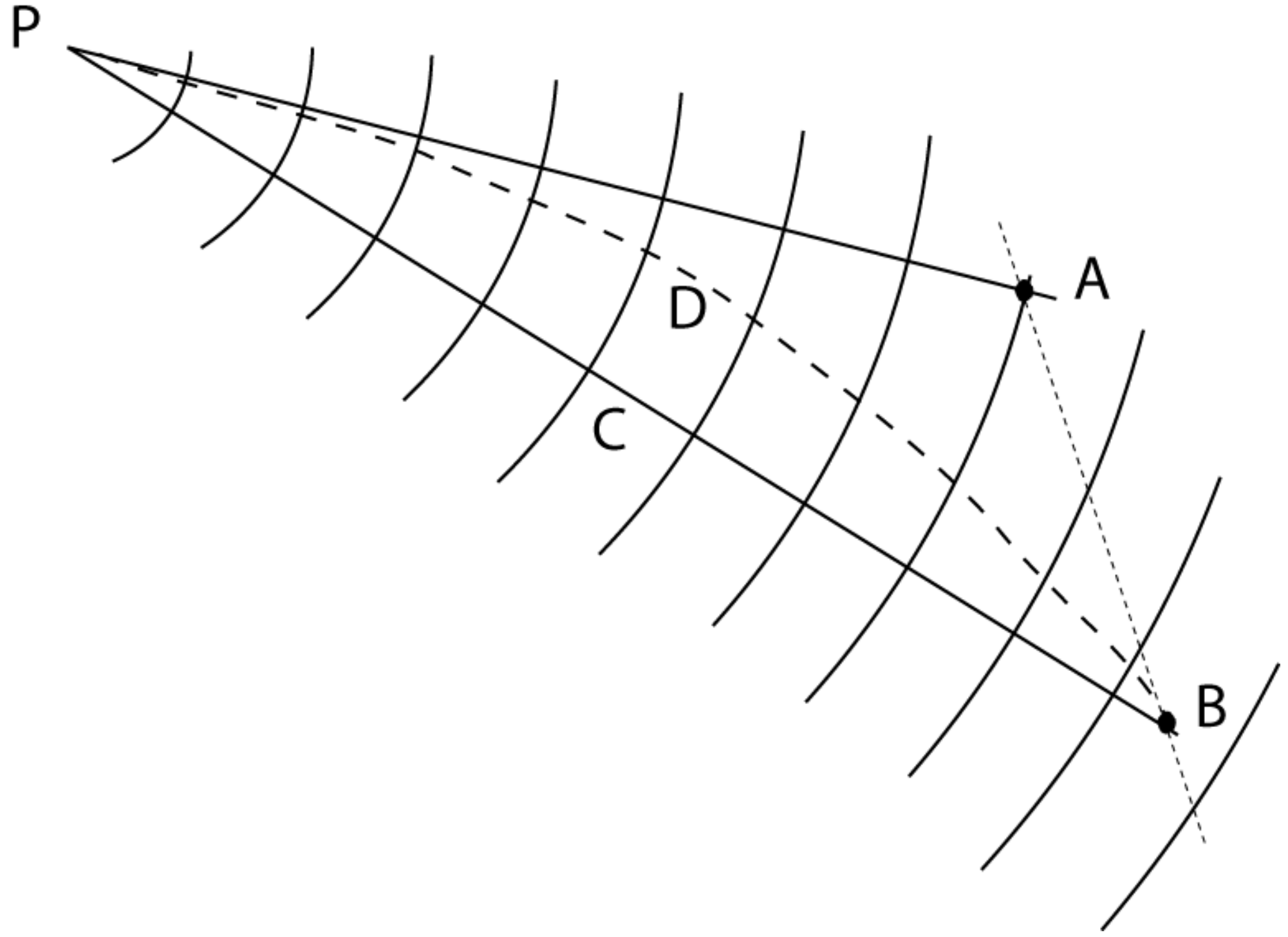

Fig. (9). Lack of Any Extra Phase Shift Due to Changing Trajectories.
A particle that would have traveled from P to A in the unperturbed beam, will be diverted to point B by a weak perturbation (along the trajectory D, denoted by the dashed line), where A and B are arbitrary points along the respective trajectories. The solid lines represent the unperturbed trajectories while the circular arcs indicate the lines of constant phase of the unperturbed wave, one wavelength apart. The line C is the unperturbed trajectory from P to B. But the length of the path of the perturbed beam D is the same as that of the unperturbed beam C, to lowest order in the perturbation. So while a different trajectory arrives at the point B, the phase will still be the same at point B as it was before the perturbation, even though a different beam arrives there. And the same is true at every point (including point A).

There is a direct, unperturbed trajectory from point P to point A. There is also a direct trajectory from P to point B, along the path C, while with the perturbation present the particle will travel the distorted path D to point B. The important feature of Fig. (9) is that to first order in the perturbation (small angle of deflection), the lengths of the paths C and D are the same. So the phase at point B has not changed even though a different wave is now arriving there. And this will be true everywhere. Although the beam is distorted, and at every arbitrary point the perturbation will ensure that a different trajectory is passing through that point, nonetheless the phase at that point will remain the same as it was before the perturbation was added. Therefore, there is no phase change at a given point due to the shifting trajectories. The only way that the phase can change at that point is if the wavelength changes, and that has already been taken into account by the first term in eq. (14).

Appendix G: Phase Shift in the Atomic Case, when Energy is Not Conserved.

In both the neutron and atom interferometer, the beam gets split at time $t_0 = 0$, the beams get re-directed at time $t_1 = T$, and recombined at time $t_2 = 2T$. In the neutron interferometer, energy is conserved, while in the atomic beam case, the upper beam (beam $u$) picks up an extra momentum and energy from the laser, $\hbar k$ and $\hbar\omega$, at time $t_0$, and loses it at time $t_1$, while the lower beam (beam $\ell$) picks it up at time $t_1$ and loses it at time $t_2$, so that the extra energy effects will drop out. In the atomic beam case the total phase accumulated ($\varphi = \frac{1}{\hbar}(\int p \cdot dr - Et)$) by the upper beam, $\varphi_u$, and the lower beam, $\varphi_\ell$, from eqs. (13) and (16), as they move along their trajectories can be seen from Fig. (10) to be

$$\varphi_u = \varphi_u(2T) - \varphi_u(0) = k_u s_u - \frac{1}{\hbar}\int_0^T U_u\,dt - (\omega_\ell + \omega)T + k_\ell s_\ell - \frac{1}{\hbar}\int_T^{2T} U_u\,dt - \omega_\ell T,$$
$$\varphi_\ell = \varphi_\ell(2T) - \varphi_\ell(0) = k_\ell s_\ell - \frac{1}{\hbar}\int_0^T U_\ell\,dt - \omega_\ell T + k_u s_u - \frac{1}{\hbar}\int_T^{2T} U_\ell\,dt - (\omega_\ell + \omega)T, \quad \text{(A22)}$$
$$\Delta\varphi \equiv \varphi_u - \varphi_\ell = -\frac{1}{\hbar}\int_0^{2T} (U_u - U_\ell)dt.$$

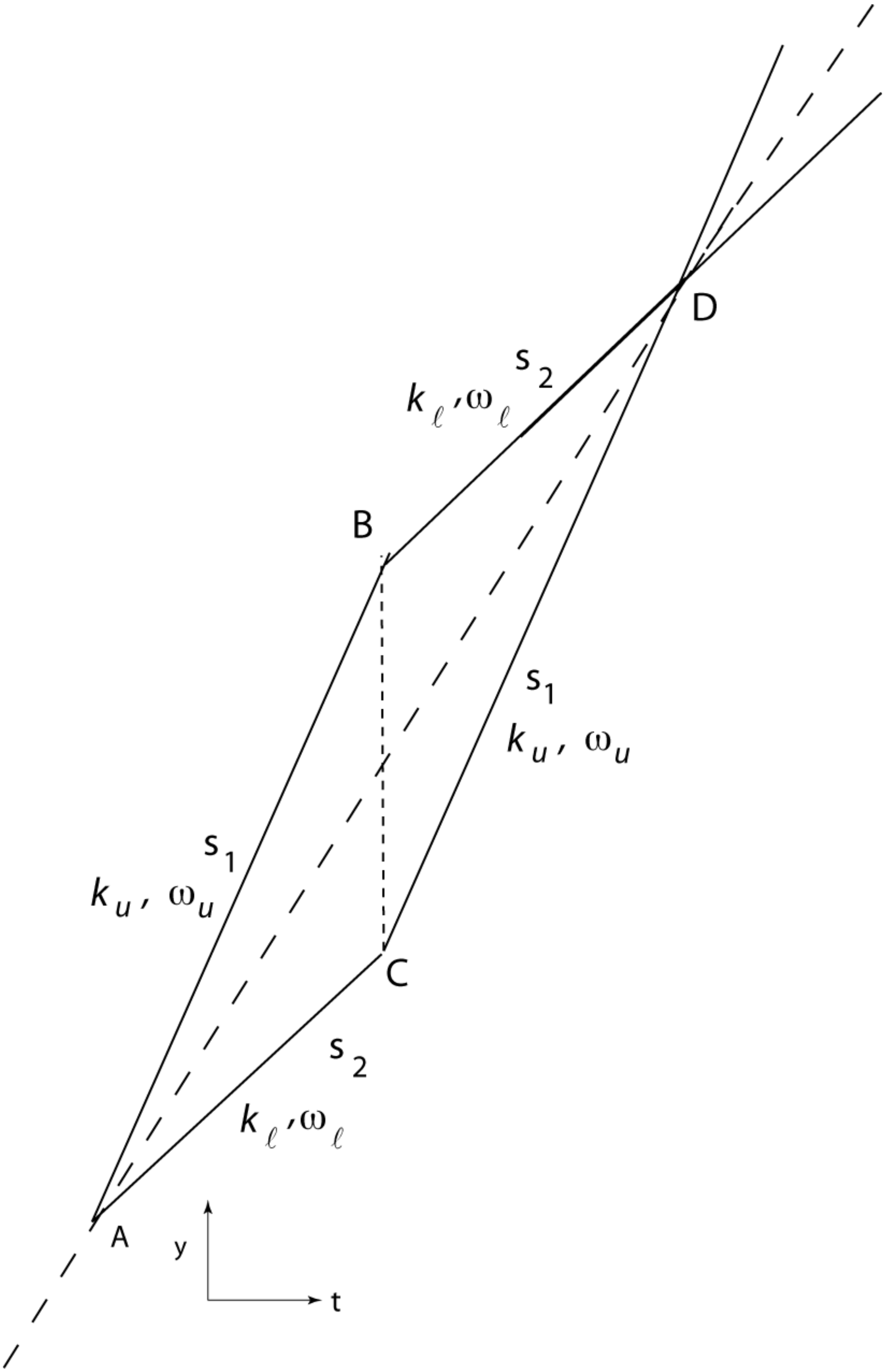

Fig. (10)

Fig. (10). Phase Shift When Energy is not Conserved. The incident atomic interferometer beam is hit by a laser at $t_0 = 0$, at point A and splits into two beams, as per Fig. (1b). The upper beam absorbs momentum $\hbar\kappa$ from the laser, so that $\hbar k_u = \hbar(k_\ell + \kappa)$, $\hbar\omega_u = \hbar(\omega_\ell + \omega)$, while the path lengths on the two segments of each beam are $s_1$ and $s_2$. Subsequently at $t_1 = T$, the upper beam loses its extra momentum, while the lower beam picks it up. Then at $t_3 = 2T$, the lower beam loses its extra momentum and the two beams recombine.

Here, $s_1$ and $s_2$ refer to the unperturbed paths of the two beams, while $\hbar k_u = \hbar(k_\ell + \kappa)$, (where $\ell, u$, refer to the lower and upper beams and $\kappa$ is the photon momentum absorbed from the laser, and similarly for the energy, $\hbar\omega$). For the neutron interferometer, the result for $\Delta\varphi$ is the same, eq. (16), since the inelastic terms from the laser do not appear, while in the atom case they cancel out. As we have indicated, the second term in eq. (14), the trajectory term, gives no contribution. (See Appendix F.)

## Appendix H: Momentum Transfer Induced by a Raman Transition

The situation corresponding to theK-C interferometer is more complicated than that described in section (3), nonetheless, eq. (A22) is correct. Rather than a photon of frequency $\omega = ck$ hitting an atom, which then absorbs the photon, the absorption takes place via a Raman transition (see Fig. (11)). There are two counter-propagating lasers of frequency $\omega_1$ and $\omega_2$, and the atom makes a hyperfine transition from state $|b\rangle$ to state $|a\rangle$ of frequency $\omega = \omega_1 - \omega_2$, by absorbing a photon at frequency $\omega_1$ and making a virtual transition to a higher P-state level of Cesium and re-emitting the photon at frequency $\omega_2$, landing in the upper hyperfine level $|a\rangle$.

This Raman technique has an important advantage. By absorbing a photon from one laser beam and emitting into the other, the momentum gain is $\hbar(|k_1|+|k_2|)$, while the energy gain is only $\hbar\omega$, which is about $10^5$ times smaller than the S-P transition energy. While this slightly complicates our argument relating to the Doppler effect, eq. (A24), it does not destroy it, but only slightly modifies it, into eq. (A27). (See Appendix I.)

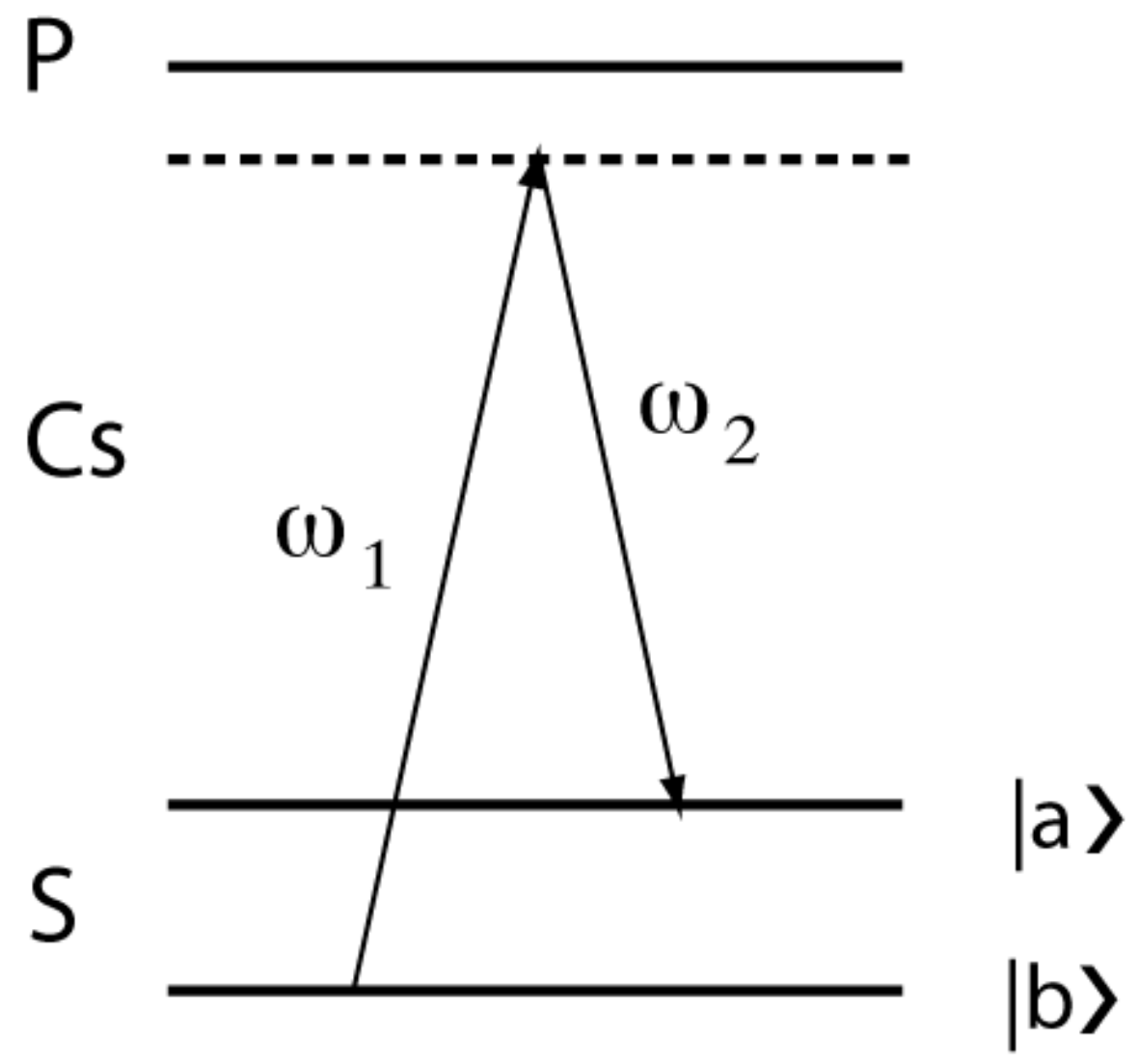

Fig. (11). Stimulated Raman Transition
In theK-C interferometer experiments[10,11], the momentum transfer to the Cesium atom was supplied by the absorption of a photon from a hyperfine level $|b\rangle$ of an S state to a virtual P state, followed by an emission of a counter-propagating photon back down to another hyperfine level $|a\rangle$ of the S state.

## Appendix I: The Changing Frequency of the Lasers and Their Effect on the Interferometer

To see how changing the frequency of the laser affects the falling atomic beam, consider the case of an atom falling in the laboratory with a constant velocity $v$. The lasers produce counter-propagating electromagnetic waves of frequency $\omega_0$, and wave numbers $k_0$ and $-k_0$. In order to keep these waves in resonance with the falling atom, they will have to change frequencies because of the Doppler effect, since the atom is moving toward the lower laser, and away from the upper one. Thus the atom sees the lower laser as having too high a frequency and the upper laser as having too low a one. So one will have to decrease the frequency of the lower laser and increase the frequency of the upper laser, in order for both to stay in resonance. This means, to first order in $v/c$, that the upper laser beam shifts from $-k_0$ to $-k_0(1 + v/c)$, while the lower beam shifts from $k_0$ to $k_0(1-v/c)$. This changes the electromagnetic potential at the virtual lattice $\Phi$, formed by the counter-propagating beams,

$$\Phi = Ae^{i(k_0 z-\omega_0 t)} + Ae^{i(-k_0 z-\omega_0 t)} = 2Ae^{-i\omega_0 t}\cos k_0 z, \qquad \text{(A23)}$$

to $\Phi'$,

$$\Phi' = Ae^{i(k_0(1-v/c)z-\omega_0(1-v/c)t)} + Ae^{i(-k_0(1+v/c)z-\omega_0(1+v/c)t)} \approx 2Ae^{-i\omega_0 t}\cos k_0(z+vt). \qquad \text{(A24)}$$

Here we have dropped a term $(\omega_0 vz/c^2)$, which is of higher order. Equation (A24) shows that for a particle moving at constant speed, shifting the frequency to stay in resonance has the effect that the lattice planes will move down at constant speed. For a slowly accelerating particle an approximate solution, good to lowest order, can be found by merely replacing $v$, by $v(t)$, and $vt$ by $\int v(t)dt$, in eq. (A24), so that the crystal planes will be accelerating by the appropriate amount. This effect is well-known and is frequently used to accelerate atoms in optical lattices (for example to create Wannier-Stark ladders[32]), and in our experiment, it cannot be ignored.

The existence of the two Raman transitions does not negate this argument. In the Raman case, if there were no Doppler effect, eq. (A23) would read

$$\begin{aligned} &\Phi = e^{ik_1 z-i\omega_1 t} + e^{i(-k_2 z-i\omega_2 t)} = 2\cos(k_+ z - \omega_- t)e^{ik_- z-i\omega_+ t}, \\ &k_\pm = (k_1 \pm k_2)/2, \quad \omega_\pm = (\omega_1 \pm \omega_2)/2. \end{aligned} \qquad \text{(A25)}$$

This equation shows that the virtual lattice is very slowly drifting with velocity $v_L = \omega_-/k_+$. If now, as before, we add a Doppler shift so that

$$\begin{aligned} &k_1 \to k_1(1-v/c), \quad \omega_1 \to \omega_1(1-v/c), \\ &-k_2 \to -k_2(1+v/c), \quad \omega_2 \to \omega_2(1+v/c), \end{aligned} \qquad \text{(A26)}$$

then

$$\Phi' \approx 2\cos\left(k_+(z+vt) - \omega_- t\right)e^{i(k_-(z+vt)-\omega_+ t)}, \qquad \text{(A27)}$$

where again we have dropped a term $(\omega_\pm vz/c^2)$, which is of higher order. This shows that as before, the lattice picks up the extra Doppler velocity. Again, if the particle is slowly accelerating, one can replace $vt$ by $\int v(t)dt$, so that the planes are slowly accelerating.

## Appendix J: Remarks on Some Discussions of the Red Shift.

Both Müller, *et al.*[23-25], and Wolf, *et al.*[26,27], have made claims as to what relativistic effects can be seen in these matter wave experiments. They have also performed extensive calculations concerning the possible equivalence-violating contributions to the experiments. In particular, Müller, *et al.*, argue that the term $mc^2/\hbar$, which has the units of a frequency, and which they call the "Compton frequency", $\omega_0$, should be considered as a basic unit of frequency reference in the problem.

It is true that rest-mass arguments enter into many phenomena, but this quantity, $\omega_0$, is in fact not the frequency of anything physical. It is a numerical invariant. If calculated in a moving system, it would still be $\omega_0$, whereas a real frequency would undergo a Doppler shift (as it is the zero-component of a four-vector). Furthermore, since the factor $e^{-i\omega_0 t}$ multiplies the wave functions of both beams, it factors out, just as the zitterbewegung does in the non-relativistic Schroedinger or Dirac equation, leaving only non-relativistic contributions, which as explained, can be seen and interpreted even in the non-relativistic limit as a proper time difference and a red shift, or more pedestrianly, as just the effect of a Newtonian potential. In the interferometer, only frequency differences between the beams are measured, and not pure frequency-like terms.

A further argument that the quantity $\omega_0$ does not enter the experiment in any significant way follows from the fact that in the energy difference between the two beams, the mass drops out. It in no way contributes to the precision of the experiment, and their claim that the precision is $\delta\omega/\omega_0$ is completely arbitrary. To see this, imagine there is another isotope of Cesium with the higher mass m', and Compton frequency ω', but with the same level structure. Then the accuracy δω would remain unchanged, but they would then claim a higher precision for the experiment of δω/ω', showing the irrelevance of the denominator ω' to the precision of the experiment.

Wolf, *et al.*, calculate $\int L_{GR}$dt for each of the two paths of the atom interferometer and find that they are equal, and conclude that matter wave interferometers cannot see differences in proper time, and so cannot measure the red shift. This conclusion is true for K-C atom interferometers, but it is not true for the COW neutron interferometer, as we have seen. They consider the situation only in the atom case, in the laboratory system, and their calculation is based on the Feynman integral, following the formulation by Storey and Cohen-Tannoudji[31]. This calculation also gives the correct the phase shift.

Wolf, *et al.*, do, however, give one incorrect argument, that the "laser-atom" part of the interaction is independent of *m*, while the proper time part is linear in *m*. Müller, *et al.*[23], as well as Wolf, *et al.*[26], point out that one can break the phase shift into 3 parts, $\varphi_{redshift}$, $\varphi_{time}$, and $\varphi_{laser}$, which they attribute to general relativity, special relativity, and the laser-atom interactions respectively, all of which contibute the same amount (to within a sign), and so they are *all* independent of *m* in their sense. This comes about because all three depend on the velocity difference between the upper and lower beams, which by momentum conservation gives $\Delta v_{at} = \Delta p_{atom}/m = \Delta p_{photon}/m = \hbar k/m$, and this $m$ cancels out the $m$ in the Lagrangian.

The definitive argument that in the neutron interferometer the proper times consumed along the two paths of the interferometer are different comes from considering the frame free-falling with the beam. In this frame there is no gravity, but the rebound of the atom from the accelerating crystal creates an asymmetrical situation between the two beams, as we have shown (see Fig. (8)). Here, each beam clearly consumes a different amount of proper time before they recombine. In the laboratory system the gravitational field (red shift) also contributes to this difference, as we have argued above. So the statement by Wolf, *et al.*, that matter wave interferometers cannot see relativistic effects must be modified to only include the K-C atom interferometer. In the neutron interferometer, the relativistic effect is present, though of course one only sees the non-relativistic residue of it, as explained. Of course the calculations of both Müller, *et al.*, and Wolf, *et al.*, are valuable concerning the non-equivalence contributions. Also these experiments[8,10,11] are still an important and elegant tool for exploring the universality of free fall (UFF).

The red shift is generally thought of as the result of a gravitational potential difference between two spatially separated points in a static metric, the effect of which is to change the rate of coordinate time at those points. The ideal way to measure it is by sending a photon from one point to the other and seeing whether it is still in resonance with the kind of atom that emitted it. Depending on how one defines ones clocks, one can attribute a lack of resonance to either a change in frequency of the photon on its trip, or a change in the absorption frequency of the atom. This kind of measurement depends on having a two-level atom that can absorb and emit photons at the two points. This type is epitomized by the famous Pound-Rebka experiment[32].

Wolf, *et al*, and Giulini[27], and especially Sinha and Samuel[33], argue that in order to measure the red shift in an interferometer experiment, one needs a two-level clock that measures time, as above, in each beam of the interferometer. But there is another way to measure this effect, and that is to make use of a proper time interval that can be integrated over the path of a particle traveling between the two points. This has been used frequently, where the creation and decay of the particle generates an interval that has been used to measure the time dilation effect[34]. In this kind of situation, the integrated time interval can record a combination of both the red shift and twin paradox effects, and so can be used to check both effects, as in the experiment of Hafele and Keating[35].

In the case of matter wave interferometers, this interval is provided by the time between when the two beams split and when they subsequently recombine to produce interference, and the difference in proper times along the two beams shows up as a phase shift produced by their interference. So in principle these interferometers detect the difference in proper times along the two paths. Since the beams travel at different heights and at different speeds, their integrated measure does sample both effects to some degree. Thus the two-level clock objection does not apply to matter wave interferometers, at least not in this type of experiment. As it happens, in the K-C atomic interferometer the effects cancel out, while in the COW neutron interferometer they add up.

# References

1. See for example J. A. Wheeler & W. H. Zurek, *Quantum Theory and Measurement*, Princeton University Press, Princeton (1983).
2. H. Rauch, W. Treimer, & U. Bonse, Test of a single crystal neutron interferometer, *Phys. Lett.* **A47**, 369-371 (1974).
3. H. Rauch and S. A. Werner, *Neutron Interferometry*, Oxford U. P., Oxford (2000).
4. R. Colella, A. W. Overhauser, & S. A. Werner, Observation of gravitationally induced quantum interference, *Phys. Rev. Lett.* **34**, 1472-1475 (1975).
5. The first atom interferometers were produced almost simultaneously in three laboratories (see also Refs. (6) and (7)): O. Carnal, and J. Mlynek, Young's double slit experiment with atoms--a simple atom interferometer, *Phys. Rev. Lett.*, **66**, 2689-2692 (1991).
6. D. W. Keith, C. R. Ekstrom, Q. A. Turchette, D. E. Pritchard, An interferometer for atoms, *Phys. Rev. Lett.* **66**, 2693-2696 (1991).
7. F. Riehle, T. Kisters, A. Witte, J. Helmcke, and C. J. Bordé, Optical Ramsey spectroscopy in a rotating frame: Sagnac effect in a matter wave interferometer, *Phys. Rev. Lett.* **67**, 177 -180 (1991).
8. The first atom interferometer using the Raman effect to create a Mach-Zehnder interferometer was by M. Kasevich and S. Chu, Atomic interferometry using stimulated Raman transitions, *Phys. Rev. Lett.* 67, 181 (1991).
9. The first atom interferometer based on light gratings was constructed by E. Rasel, *et al.*, Atom wave interferometry with diffraction gratings of light, *Phys. Rev. Lett.* **75**, 2633-2637 (1995).
10. A. Peters, K. Y.Chung, & S. Chu, High-precision measurement using atom interferometry, *Metrologia* **38**, 25-61 (2001).
11. A. Peters, K. Y. Chung, & S. Chu, A measurement of gravitational acceleration by dropping atoms, *Nature* **400**,8849-852 (1999).
12. W. Schleich, M. O. & Scully, in: New Trends in Atomic Physics, *Proceedings of the Les Houches Summer School, Session XXXVIII,* 1982, R. Stora, & G. Grynberg, editors, North Holland, Amsterdam, (1984).
13. T. Van Zoest, *et al*, Bose-Einstein-Condensation in micro-gravity, Science **328**, 1540-1543 (2010), doi:10.1126/science.1189164; E. Arimondo, W. Ertmer, E. M. Rasel, & W. P. Schleich, editors, "Atomic and Space Physics", *Proceedings of the International School of Physics "Enrico Fermi"*, Elsevier, Amsterdam (2009).

14. V. V. Nesvizhevsky, *et al*, Quantum states of neutrons in the earth's gravitational field, *Nature* **415**, 297-299 (2002).
15. H. Abele, T. Jenke, D.Stadler, & P. Geltenbort, QuBounce: the dynamics of ultra-cold neutrons falling in the gravity potential of the earth, *Nuclear Physics* **A827**, 593c-595c, (2009), doi:10.1016/ *J. Nucl. Phys A* 2009.05.131.
16. H. Abele, T. Jenke, H. Leeb, & J. Schmiedmayer, Ramsey's method of separated oscillating fields and its application to gravitationally induced quantum shift, *Phys. Rev.* ***D*81**, 065019 (2010), doi:10.1038/415297a.
17. I. Ciufolini, & E. C. Pavils, A confirmation of the general relativistic prediction of the Lense-Thirring effect, Nature, 431, 958-960, 2004. This claim has proven to be controversial.
18. See, *e.g*., L. Iorio, , A critical analysis of a recent test of the Lense-Thirring effect with the LAGEOS satellites, Journal of Geodesy 80, 123-136 (2006).
19. A recent direct test has been completed, C. W. F. Everitt, *et al*., Gravity Probe B: Final Results of a Space Experiment to Test General Relativity, arXiv:1105.3456v1 [gr-qc] 17 May 2011; *Phys. Rev. Letters* **106**, 221101 (2011).
20. D. M. Greenberger, The neutron interferometer as a device for illustrating the strange behavior of quantum systems, *Rev. Mod. Physics* **55**, 875-905 (1983).
21. D. M. Greenberger, & A. W. Overhauser, Coherence effects in neutron diffraction and gravity experiments, *Rev. Mod. Physics* **51**, 43-78 (1979).
22. C. M. Will, The confrontation between general relativity and experiment, *Living Rev. Relativity* **9**, 3 (2006), http://www.livingreviews.org/Irr-2006-3.
23. H. Müller, A. Peters, & S. Chu, A precision measurement of the gravitational redshift by the interference of matter waves, *Nature* **463**, 926-929 (2010).
24. H. Müller, *et al*., *Nature* doi: 10.1038/ *Nature* 09341 (2010).
25. M. A. Hohensee, S. Chu, A. Peters, and H. Müller, Equivalence principle and gravitational redshift, Phys. Rev. Letters **106**, 151102 (2011); arXiv: 1102.4361v1 [gr-qc]21 Feb. 2011.
26. P. Wolf, *et al*., *Nature* doi: 10.1038/ *Nature* 09340 (2010).
27. P. Wolf, *et al*., Does an atom interferometer test the gravitational red shift at the Compton frequency?, *Class. and Qu. Gravity* **28**, 145017 (2011); doi: 10.1088//0264-9381/28/14/145017; *arXiv*:1012.1194v1 [gr-qc] 6 Dec (2010).
28. D. Giulini, Equivalence Principle, Quantum Mechanics, and Atom-Interferometric Tests, Proceedings of the Conference *Quantum Field Theory and Gravity*, F. Finster, *et al*., Eds., Birkhäuser, Basel (2012).
29. D. M. Greenberger, Inadequacy of the usual Galilean transformation in quantum mechanics, *Phys. Rev. Lett.* **87**(10), 100405-1 (2001)

30. E. Kajari, N. L. Harshman, E. M. Rasel, S. Stenholm, G. Süssmann, & W. P. Schleich, Inertial and gravitational mass in quantum mechanics, *App. Phys.* **B100**,43-60 (2010)/doi 10.1007/s00340-010-4085-8.
31. W. P. Schleich, D. M. Greenberger, and E. M. Rasel, A representation-free description of the Kasevich-Chu interferometer: a resolution of the redshift controversy, (to be published in the New Journal of Physics).
32. A translation of De Broglie's thesis appears in J. W. Haslett, translation of L. de Broglie, Recherches sur la théorie des Quanta, University of Paris (1924), *Am. J. Phys.* **40,** (1972).
33. J-L. Staudenmann, S. A. Werner, R. Colella, and A. W. Overhauser, Gravity and inertia in quantum mechanics, Phys. Rev. **A21**, 1419-1438 (1980).
34. M. Kasevich and S. Chu, Measurement of the gravitational acceleration of an atom with a light-pulse atom interferometer, Appl. Phys. **B54**, 321-332 (1992).
35. P. Storey, and C. Cohen-Tannoudji, The Feynman path integral approach to atomic interferometry. A tutorial, *J. Phys. II France* **4**, 1999-2027 (1994).
36. S. R. Wilkinson, *et al.*, Observation of Wannier-Stark ladders in accelerating optical potential, *Phys. Rev. Lett.* **76**, 4512-15 (1996).
37. R. V. Pound, and G. A. Rebka, Apparent Weight of Photons, Phys. Rev. Letters **4**, 337-341 (1960).
38. S. Sinha, and J. Samuel, Atom Interferometry and the Gravitational Redshift, *Class. and Qu. Gravity*, **28**, 145018 (2011)/doi: 10.1088/0264-9381/28/14/145018; arXiv:1102.2587v4 [gr-qc] 16 May 2011.
39. One of the earliest is B. Rossi, and D. B. Hall, Variations of the rate of decay of mesotrons with momentum, *Phys. Rev.* **59**, 223-228 (1941).
40. J. Hafele and R. Keating, (July 14, 1972). Around the world atomic clocks predictedrelativistic time gains, *Science* **177** (4044): 166–168, 1972